\documentclass[         
aps,                    %  American Physical Society
prd,                    %  Physical Review D
showpacs,               %  Displays PACS after abstract
nofootinbib,            %  Does not treat footnotes as references
showkeys,               %
preprintnumbers,        %
floatfix]               %  Fixes float errors
{revtex4}               %  REVTEX 4 Package used
\usepackage{graphicx,longtable}
\begin{document}
%-----------------------------------------------------------------%
\title{Possible CP-violation effects in core-collapse Supernovae}

\author{        A.~B. Balantekin}
\email{         baha@physics.wisc.edu}
%-----------------------------------------------------------------%
\affiliation{  Department of Physics, University of Wisconsin\\
               Madison, Wisconsin 53706 USA }
%-----------------------------------------------------------------%
\author{J. Gava}
\email{gava@ipno.in2p3.fr}
\author{        C. Volpe}
\email{volpe@ipno.in2p3.fr}
\affiliation{Institut de Physique Nucl\'eaire, F-91406 Orsay cedex,
France}

\date{\today}
%-----------------------------------------------------------------%
\begin{abstract}
We study CP-violation effects when neutrinos are present in dense matter,
such as outside the proto-neutron star formed in a core-collapse 
supernova. Using general arguments based on the Standard Model, we 
confirm that there are no CP-violating effects at the tree level on the 
electron neutrino and anti-neutrino fluxes in a core-collapse supernova. 
On the other hand significant effects can be obtained for muon and 
tau neutrinos even at the tree level. 
We show that CP violating effects can be present in the supernova electron 
(anti)neutrino fluxes as well, if muon and tau neutrinos have different 
fluxes at the neutrinosphere. Such differences could arise due to physics 
beyond the Standard Model, such as the presence of flavor-changing 
interactions. 
\end{abstract}
%-----------------------------------------------------------------%
\medskip
\pacs{}
\keywords{CP-violation in neutrino sector, r-process nucleosynthesis, 
core-collapse supernova}
\preprint{}
\maketitle
%------------------------------------------------------------------%
                                                      
%-----------------------------------------------------------------%
\section{Introduction}

Recent results from solar, atmospheric and reactor experiments 
have significantly improved our knowledge of the neutrino mass differences 
and of  two of the 
mixing angles. If the remaining mixing angle, $\theta_{13}$, is relatively 
large there is a possibility that violation of CP symmetry may be observable 
in the neutrino sector. Currently planned and future experiments will have 
improved
sensitivities to the value of this angle (see e.g. 
\cite{Guo:2007ug,Ardellier:2006mn,Kato:2007zz}).
Effects of CP-violation in accelerator neutrino 
oscillation experiments have been extensively investigated 
\cite{Dick:1999ed,Peres:2003wd,Ishitsuka:2005qi,Barger:1980jm,Karagiorgi:2006jf,Volpe:2006in}. 
The discovery of a non zero Dirac delta phase might help our 
understanding of the observed matter-antimatter 
asymmetry of the universe 
\cite{Pascoli:2006ie,Pascoli:2006ci,Mohapatra:2006se,Barger:2003gt}. 
Besides studies on terrestrial experiments with man-made sources, 
a few recent works have addressed
CP-violation with neutrinos from astrophysical sources 
(see e.g. \cite{Akhmedov:2002zj,Winter:2006ce}). 
The purpose of 
the present paper is to explore possible 
effects coming from the CP-violating phase in dense matter, 
such as that encountered in core-collapse supernovae. 

Core-collapse supernovae occur following the stages of nuclear burning 
during stellar evolution after an iron core is formed.  
The iron cores formed during the evolution of massive stars are supported 
by the electron degeneracy pressure and hence are unstable against a collapse 
during which most of the matter is neutronized. Once the density exceeds 
the nuclear density this 
collapse is halted. Rebounding pressure waves break out 
into a shock wave near the sonic point where the density reaches the nuclear 
density. Evolution of this shock wave and whether it produces an explosion 
is a point of current investigations. However, it is observed that the 
newly-formed hot proto-neutron star cools by neutrino emission. 
Essentially the entire 
gravitational binding energy of eight or more solar mass star is radiated away 
in neutrinos. Although the initial collapse is a very orderly (i.e. low 
entropy) process, during the cooling stage 
at later times a neutrino-driven wind heats the neutron-rich 
material to high entropies 
\cite{Woosley:1992ek,Woosley:1994ux,Takahashi:1994yz}. 

Neutrino interactions play a very important role in the evolution of 
core-collapse supernovae and in determining the element abundance 
\cite{Balantekin:2003ip}. Neutrino heating is a 
possible mechanism for reheating the stalled shock \cite{Bethe:1984ux}. 
A good fraction of the heavier nuclei were formed in the rapid neutron 
capture (r-process) nucleosynthesis scenario \cite{Burbidge:1957vc}.
Core-collapse supernovae are one of the possible sites for the r-process 
nucleosynthesis. A key quantity 
for determining the r-process yields is the neutron-to-seed nucleus ratio, 
determined by the neutron-to-proton ratio, which is controlled 
by the neutrino fluxes. 
 In addition, recent work indicates
that neutrino-neutrino interactions plays a potentially very significant
role in supernovae
\cite{Samuel:1993uw,Sigl:1992fn,Balantekin:2004ug,Duan:2006jv,Balantekin:2006tg,Hannestad:2006nj}.

In this paper we study CP violation aspects in the core-collapse 
supernova environment.
We first analyze analytically and in general terms, how the 
neutrino propagation equations and the
evolution operator are modified in matter, in presence of a 
non-zero Dirac delta
phase. We obtain a general formula which is valid for any
matter density profile\footnote{Such findings are in agreement with what 
was found in
Ref. \cite{Yokomakura:2002av}.}. In particular we demonstrate that, as in 
vacuum, the electron (anti)neutrino
fluxes are independent of the phase $\delta$, if mu and tau neutrinos have
the same fluxes at the neutrinosphere in the supernova\footnote{A remark 
on this
aspect was made in \cite{Yoshida:2006sk}.}. On the other hand the electron 
(anti)neutrino fluxes will depend on $\delta$, 
if mu and tau neutrinos have different fluxes at the neutrinosphere, at 
variance with what was found in \cite{Akhmedov:2002zj}. We present  
numerical calculations
on possible CP violation effects on the mu and tau neutrino fluxes as 
well as on the 
electron (anti-)neutrino flux and the electron fraction. The latter can 
only appear if
physics beyond the Standard Model, such as flavor changing interactions, 
induces differences on the mu and tau neutrino initial total 
luminosities and/or temperatures. Finally we calculate the effects from the CP violating phase 
on the number of events in an observatory on earth.

The plan of this paper is as follows. In Section II we present the general
formalism to describe the neutrino evolution in presence of the 
$\delta$ phase. The formulas concerning neutrino fluxes and the electron 
fraction in the
supernova environment are recalled in Section III. Numerical results on
these quantities are presented in Section IV. Conclusions are made in
Section V.

\section{Neutrino mixing in the presence of CP-violating phases}
\subsection{Neutrino mixing  in ordinary matter in presence of 
CP-violating phases}
The neutrino mixing matrix is $U_{\alpha i}$ where $\alpha$ denotes the
flavor index and $i$ denotes the mass index:
\begin{equation}
  \label{udef}
  \Psi_{\alpha} = \sum_i U_{\alpha i} \Psi_i .
\end{equation}
For three neutrinos we take
\begin{equation}
\label{mixing}
U_{\alpha i} = T_{23}T_{13}T_{12} = \left(\matrix{
     1 & 0 & 0  \cr
     0 &  C_{23}  & S_{23} \cr
     0 & - S_{23} &  C_{23} }\right)
 \left(\matrix{
     C_{13} & 0 &  S_{13} e^{-i\delta}\cr
     0 &  1 & 0 \cr
     - S_{13} e^{i\delta} & 0&  C_{13} }\right)
 \left(\matrix{
     C_{12} & S_{12} &0 \cr
     - S_{12} & C_{12} & 0 \cr
     0 & 0&  1 }\right) ,
\end{equation}
where $C_{13}$, etc. is the short-hand notation for $\cos
{\theta_{13}}$, etc. and $\delta$ is the CP-violating phase. 
The MSW equation is 
\begin{equation}
\label{msw}
i \frac{\partial}{\partial t} \left(\matrix{ \Psi_e\cr \Psi_{\mu} \cr
    \Psi_{\tau}} \right)  = \left[ T_{23}T_{13}T_{12} \left(\matrix{
     E_1 & 0 & 0  \cr
     0 &  E_2  & 0 \cr
     0 & 0 &  E_3 }\right)
T_{12}^{\dagger}T_{13}^{\dagger} T_{23}^{\dagger} +
\left(\matrix{
     V_c+V_n & 0 & 0  \cr
     0 &  V_n  & 0 \cr
     0 & 0 &  V_n }\right)\right] \left(\matrix{ \Psi_e\cr \Psi_{\mu}
    \cr     \Psi_{\tau}} \right),
\end{equation}
where 
\begin{equation}
  \label{wolfen1}
V_c (x) = \sqrt{2} G_F  N_e (x)
\end{equation}
for the charged-current and
\begin{equation}
  \label{wolfen2}
V_n (x) =  - \frac{1}{\sqrt{2}} G_F N_n(x).
\end{equation}
for the neutral current. Since $V_n$ only contributes an overall phase
to the neutrino evolution 
we ignore it\footnote{
The results we show in the present work do not include the difference in the 
mu and tau refractive indices which appear at one loop level due to 
different muon and tau lepton masses, $V_{\mu \tau}$ \cite{Botella:1986wy}, 
which is $10^{-5}~ V_c$. In fact, we have tested that the inclusion of this
correction modifies very little the numerical results presented in Section III. }. 
Following references 
\cite{Balantekin:1999dx} and \cite{Balantekin:2003dc}
we introduce the combinations
\begin{eqnarray}
\label{rot1}
\tilde{\Psi}_{\mu} &=& \cos{\theta_{23}} \Psi_{\mu} -
\sin{\theta_{23}} \Psi_{\tau}, \\
\tilde{\Psi}_{\tau} &=& \sin{\theta_{23}} \Psi_{\mu} +
\cos{\theta_{23}} \Psi_{\tau},
\end{eqnarray}
which corresponds to multiplying the neutrino column vector in
Eq. (\ref{msw}) with $T_{23}^{\dagger}$ from the left. Eq. (\ref{msw})
then becomes
\begin{equation}
\label{mswmod1}
i \frac{\partial}{\partial t} \left(\matrix{ \Psi_e\cr
    \tilde{\Psi}_{\mu} \cr \tilde{\Psi}_{\tau} }\right)
= \left[       T_{13}T_{12} \left(\matrix{
     E_1 & 0 & 0  \cr
     0 &  E_2  & 0 \cr
     0 & 0 &  E_3 }\right)
T_{12}^{\dagger}T_{13}^{\dagger}                  +
\left(\matrix{
     V_c   & 0 & 0  \cr
     0 &  0    & 0 \cr
     0 & 0 &  0   }\right)\right] \left(\matrix{ \Psi_e\cr
    \tilde{\Psi}_{\mu}  \cr     \tilde{\Psi}_{\tau} }\right) .
\end{equation}
We define 
\begin{equation}
\label{ham}
\tilde{H} 
= \left[       T_{13}T_{12} \left(\matrix{
     E_1 & 0 & 0  \cr
     0 &  E_2  & 0 \cr
     0 & 0 &  E_3 }\right)
T_{12}^{\dagger}T_{13}^{\dagger}                  +
\left(\matrix{
     V_c   & 0 & 0  \cr
     0 &  0    & 0 \cr
     0 & 0 &  0   }\right)\right]. 
\end{equation}
The Hamiltonian $\tilde{H}$ depends on the CP-violating phase, $\delta$. 
It is lengthy but straightforward to show that
\begin{equation}
\label{hamred}
\tilde{H}(\delta) = S^{\dagger} \tilde{H}(\delta=0) S,
\end{equation}
where
\begin{equation}
\label{s}
S^{\dagger} = \left(\matrix{
     1   & 0 & 0  \cr
     0 &  1    & 0 \cr
     0 & 0 &  e^{i\delta}   }\right) .
\end{equation}
We are interested in solving the evolution equation corresponding to Eq. 
(\ref{mswmod1}):
\begin{equation}
\label{ev}
i \hbar \frac{d\hat{U}}{dt} = \tilde{H} \hat{U}. 
\end{equation}
It is important to recall that we need to solve this equation with the 
initial condition 
\begin{equation}
\label{inconu}
\hat{U}(t=0)=1.
\end{equation} 
Defining 
\begin{equation}
\label{us}
{\cal U}_0 = S \hat{U},
\end{equation} 
and using the relation in Eq. (\ref{hamred}) we get
\begin{equation}
\label{ev1}
i \hbar \frac{d{\cal U}_0}{dt} = \tilde{H}(\delta =0) \> {\cal U}_0,
\end{equation}
i.e., ${\cal U}_0$ provides the evolution when the CP-violating phase is 
set to zero. Using Eq. (\ref{inconu}) we see that the correct initial 
condition on ${\cal U}_0$ is ${\cal U}_0(t=0)=S$. However, Eq. (\ref{ev1}) 
is nothing but the neutrino evolution equation with CP-violating phase set 
equal to zero. If we call the solution of this equation with the standard 
initial condition $\hat{U}_0(t=0)=1$ to be $\hat{U}_0$, 
we see that we should set 
${\cal U}_0 = \hat{U}_0 S$, which yields
\begin{equation}
\label{main}
\hat{U}(\delta) = S^{\dagger} \hat{U}_0 S. 
\end{equation}
Eq. (\ref{main}) illustrates how the effects of the CP-violating phase 
separate in describing the 
neutrino evolution. 
It is valid both in vacuum and in matter. This result is in agreement with 
\cite{Yokomakura:2002av}.
It is easy to verify that this result does not depend 
on the choice of the parametrization for the neutrino mixing matrix. 

Using Eq. (\ref{main}) it is possible to relate survival probabilities 
for the two cases with $\delta = 0$ and $\delta \neq 0$. We define the 
amplitude for the process $\nu_x \rightarrow \nu_y$ to be $A_{xy}$ when 
$\delta \neq 0$ and to be $B_{xy}$ when $\delta =0$ so that 
\begin{equation}
P(\nu_x \rightarrow \nu_y, \delta \neq 0) = |A_{xy}|^2.
\end{equation}
and 
\begin{equation}
P(\nu_x \rightarrow \nu_y, \delta =0) = |B_{xy}|^2.
\end{equation}
Using Eq. (\ref{main}) one can immediately see that the electron neutrino 
survival probability does not depend on the CP-violating phase. 
One can further write  
\[
c_{23} A_{\mu e} - s_{23} A_{\tau e} = c_{23} B_{\mu e} - s_{23} B_{\tau e}
\]
\[
s_{23} A_{\mu e} + c_{23} A_{\tau e} = e^{-i\delta} 
[ s_{23} B_{\mu e} + c_{23} B_{\tau e}]
\]
By solving these equations one gets 
\begin{equation}
\label{mue}
A_{\mu e} = (c_{23}^2 + s_{23}^2 e^{- i \delta}) B_{\mu e} + c_{23}s_{23} 
(e^{- i \delta} -1) B_{\tau e},
\end{equation}
and
\begin{equation}
\label{taue}
A_{\tau e} = c_{23}s_{23} 
(e^{- i \delta} -1) B_{\mu e} + (s_{23}^2 + c_{23}^2 e^{- i \delta})
B_{\tau e}. 
\end{equation}
Clearly the individual amplitudes in Eqs. (\ref{mue}) and (\ref{taue}) 
depend on the CP-violating phase. However, 
taking absolute value squares of 
Eqs. (\ref{mue}) and (\ref{taue}), after some algebra one obtains that
\begin{equation}
\label{sq}
|A_{\mu e}|^2 + |A_{\tau e}|^2 = |B_{\mu e}|^2 + |B_{\tau e}|^2,
\end{equation} 
or equivalently
\begin{equation}
\label{sq2}
P (\nu_{\mu} \rightarrow \nu_e, \delta \neq 0) + 
P (\nu_{\tau} \rightarrow \nu_e, \delta \neq 0) =
P (\nu_{\mu} \rightarrow \nu_e, \delta =0) + 
P (\nu_{\tau} \rightarrow \nu_e, \delta =0) .
\end{equation} 

Since $P(\nu_e \rightarrow \nu_e)$ does not depend on $\delta$, 
Eq. (\ref{sq2}) implies that if one starts with identical spectra with tau and 
mu neutrinos, one gets the same electron neutrino spectra no matter what the 
value of the CP-violating phase is (also see Eq.\ref{e:flux}). 
This result was first established in \cite{Akhmedov:2002zj} 
with a different derivation.
A remark on this aspect is also made in \cite{Yoshida:2006sk}.

Some differences in the muon/tau neutrino fluxes at emission can arise
at the level of the Standard Model, from example from radiative corrections
to the muon and tau neutrino cross sections \cite{Botella:1986wy}. 
On the other hand, if physics beyond the Standard model 
operates during the infall and the shock-bounce stages of the supernova 
evolution, mu and tau neutrino fluxes can differ and induce CP-violating 
effects in the supernova environment.  For example, generic 
neutrino-flavor changing interactions 
can give rise to significant net mu and tau lepton 
numbers \cite{Amanik:2004vm}. In particular, if there are flavor 
changing interactions involving charged leptons (e.g. a large scale 
conversion in the $e^- \rightarrow \mu^-$ channel) one could also end up 
with significantly 
different mu and tau neutrino fluxes. In such cases one could 
have effects from the CP-violating phase on the electron (anti)neutrino 
fluxes as well.

Note that our findings are at variance with those of Ref.\cite{Akhmedov:2002zj}. 
In fact, there the authors conclude that even if mu and tau neutrino fluxes are
different, CP-violation effects cannot be observed. Such a difference arises 
from the fact that different initial conditions are taken in our calculations
compared to those used in 
Eq.(47) of \cite{Akhmedov:2002zj}. Indeed since
the initial neutrino states should be those at the neutrinosphere, 
the neutrino
conversion probability $P_{i\alpha}$ should depend on the $\delta$ phase
(see section 3.4 of \cite{Akhmedov:2002zj}).

\section{Neutrino Fluxes and Electron Fraction in Supernovae}

In this work we will discuss possible effects induced by the 
Dirac CP violating phase
on two particular observables in the core-collapse supernova environment:
the neutrinos fluxes $\phi_{\nu}$ and the electron fraction $Y_{e}$. 
Note that the impact of the neutrino magnetic moment on 
such observables was studied in
\cite{Balantekin:2007xq}.
According to supernova simulations,
the neutrino fluxes at the neutrinosphere
are quite well described by Fermi-Dirac distributions \cite{Raffelt:1996wa}
or power-law spectra \cite{Keil:2002in}. Neutrino masses and mixings 
modify this simple pattern by mixing the spectra during neutrino evolution.
Since muon and tau neutrinos only undergo neutral current interactions,
they decouple deeper in the star. Electron (anti-)neutrinos experience
both charged and neutral current interactions, the anti-neutrino
cross sections being weaker than for neutrinos and matter being 
neutron-rich. As a result
a neutrino hierarchy of temperatures is expected, 
$\langle E_{\nu_e} \rangle < \langle E_{\bar{\nu}_{e}} \rangle 
< \langle E_{\nu_{\tau}} \rangle$ with typical ranges of 10-13,  13-18 and 
18-23 MeV respectively \cite{Keil:2002in}.
In order to show possible CP-violating effects on the 
$\nu_{i}$ fluxes, we will use the ratio:
\begin{equation}\label{e:ratio}
R_{\nu_i}(\delta) = {{\phi}_{\nu_i}(\delta) \over{{\phi}_{\nu_{i}}(\delta=0^{\circ})}}
\end{equation}
where the neutrino fluxes are given by
\begin{equation}\label{e:flux}
{\phi}_{\nu_i}(\delta) =  L_{\nu_i}P(\nu_i \rightarrow \nu_i) + 
L_{\nu_j}P(\nu_j \rightarrow \nu_i)+L_{\nu_k}P(\nu_k \rightarrow \nu_i)
\end{equation}
with the luminosities 
\begin{equation}\label{e:lum}
L_{\nu_i}(r,E_{\nu})= {L^0_{\nu_i} \over{4 \pi r^2 (kT)^3 \langle E_{\nu} \rangle F_2(\eta)}}{{E_{\nu}^2} \over{1 + \exp{(E_{\nu}/T_{\nu} - \eta)}}}
\end{equation}
where $F_2(\eta)$ is the Fermi integral,
$L_0$ is the luminosity that we take as $6 \times 10^{51}$erg/s as an example and 
and $r$ is the distance
from the proto-neutron star.
We consider the Fermi-Dirac distribution as typical
example. The quantities $P(\nu_i \rightarrow \nu_{i/j})$ correspond to
the survival/appearance neutrino probability during the evolution in
matter.

The dominant reactions that control the neutron-to-proton 
ratio outside the hot proto-neutron star is the capture
reactions on free nucleons
\begin{equation}
\label{rate1}
\nu_e + {\rm n}  \rightleftharpoons {\rm p}+ e^{-} ,
\end{equation}
and
\begin{equation}
\label{rate2}
\bar{\nu}_e + {\rm p} \rightleftharpoons {\rm n} + e^{+} .
\end{equation} 
We designate the rates of the
forward and backward reactions in Eq. (\ref{rate1}) to be 
 $\lambda_{\nu_e}$ and $\lambda_{e^-}$ and 
the rates of the
forward and backward reactions in Eq. (\ref{rate2}) to be
 $\lambda_{\bar{\nu}_e}$ and $\lambda_{e^+}$. 
The electron fraction, $Y_e$, is the net number of electrons (number
of electrons minus the number of positrons) per baryon:
\begin{equation}
\label{e:Ye}
Y_e = ( n_{e^-} - n_{e^+} ) / n_B ,
\end{equation}
where $n_{e^-}$, $n_{e^+}$, and $n_B$ are number densities of
electrons, positrons, and baryons, respectively.
If no heavy nuclei are present we can write the rate of change of 
$Y_e$ as
\begin{equation}
\label{Yerate}
\frac{dY_e}{dt} = \lambda_n - ( \lambda_p + \lambda_n) Y_e
+ \frac{1}{2} ( \lambda_p - \lambda_n ) X_{\alpha},
\end{equation}
where we introduced the alpha fraction $X_{\alpha}$, 
the total proton loss rate $\lambda_p =
\lambda_{\bar{\nu}_e} + \lambda_{e^-}$ and the total neutron loss
rate $\lambda_n = \lambda_{\nu_e} + \lambda_{e^+}$. 
From Eq. (\ref{Yerate}) one can write the equilibrium value of the
electron fraction
\begin{equation}
\label{Yeeq}
Y_e = \frac{\lambda_n}{\lambda_p + \lambda_n}
+ \frac{1}{2} \frac{\lambda_p - \lambda_n}{\lambda_p + \lambda_n}
X_{\alpha} .
\end{equation}
As the alpha particle mass fraction increases more and more  free
nucleons get bound in alpha particles \cite{Fuller:1995ih}.  This
phenomenon, called alpha  effect, pushes the electron fraction towards
towards the value 0.5 (cf.  Eq. (\ref{Yeeq})). Since it reduces
available free neutrons, alpha effect  is a big impediment to
r-process nucleosynthesis \cite{Meyer:1998sn}.  At high temperatures,
alpha particles are absent and the second term in Eq.(\ref{Yeeq})  can
be omitted. Since electron and positron capture rates are very small,
the electron fraction can be rewritten as
\begin{equation}
Y_e^{(0)} = \frac{1}{1 + \lambda_p/\lambda_n}
\label{e:Yeeq0}
\end{equation}
with the capture rates on $x=p,n$ given by
\begin{equation}
\lambda_{n,p} = \int \sigma_{\nu_e n,\bar{\nu}_e p} 
(E_{\nu}) \phi(E_{\nu})dE_{\nu}
\label{e:capture}
\end{equation}
and $\sigma_{\nu_e n,\bar{\nu}_e p}$ being the reaction cross sections 
for the corresponding processes 
Eqs.(\ref{rate1}-\ref{rate2}).

\section{Possible CP violation effects: Numerical results}
\noindent
It is the goal of this section to investigate numerically 
effects induced by the Dirac $\delta$ phase :
i) on the muon and tau neutrino fluxes when their fluxes at the 
neutrinosphere are supposed to be 
equal; ii) on the electron, muon, tau (anti)neutrino fluxes, when
the muon and neutrino fluxes differ at the neutrinosphere. 
In fact, Eqs.(\ref{udef}-\ref{sq2}) 
and (\ref{e:flux}) show that in the latter case the electron 
(anti)neutrino fluxes become
sensitive to the CP violating phase.  
We have performed calculations for several 
values of the phase. The effects discussed here are present for
any value and maximal for $\delta = 180^\circ$. For this reason
most of the numerical results we show correspond to this value.  

\begin{figure}[t]
\vspace{.6cm}
\includegraphics[scale=0.3,angle=0]{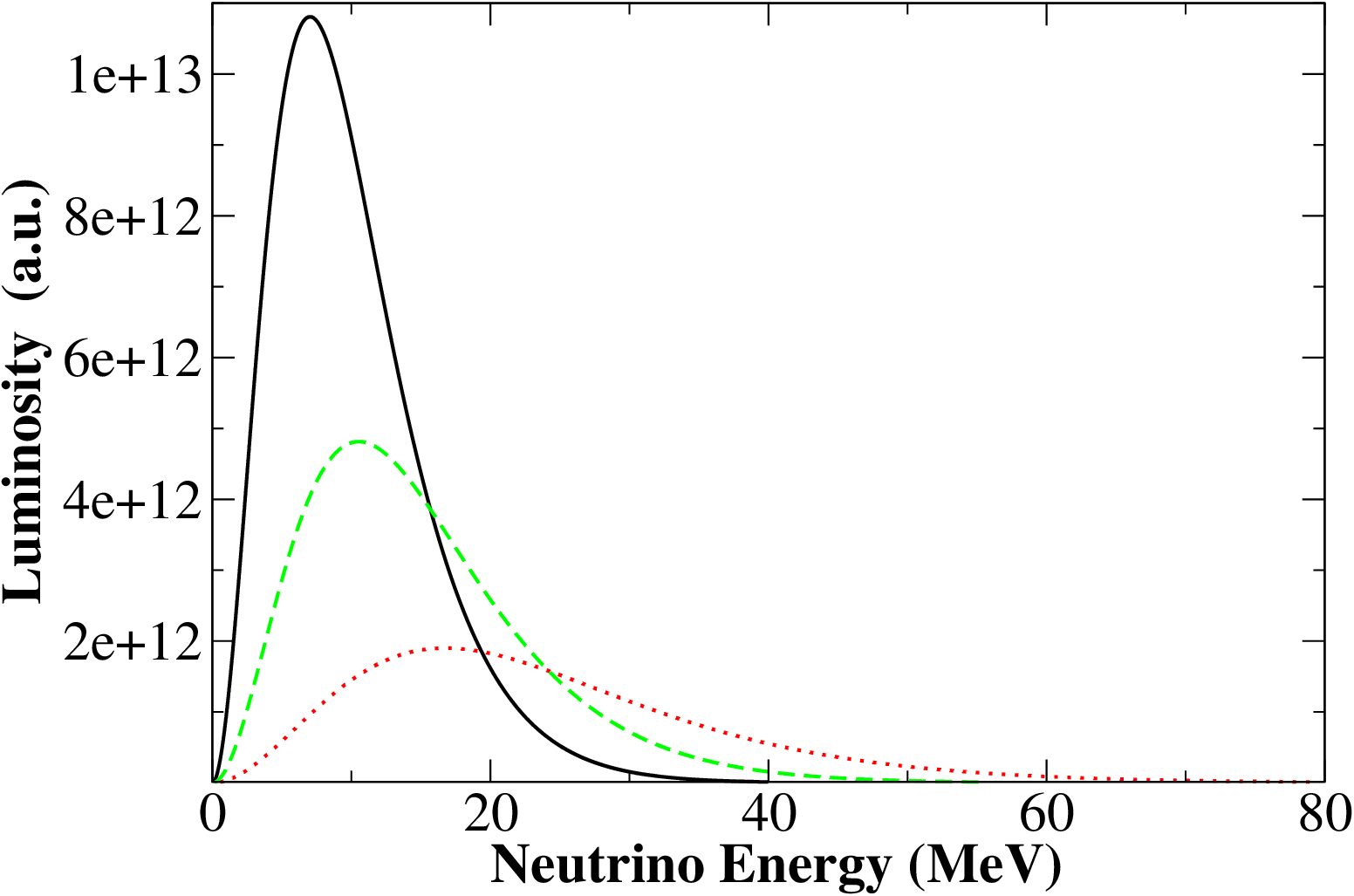}
\caption{Neutrino fluxes at the neutrinosphere: the curves show the
Fermi-Dirac distributions used for electron neutrinos with
T$_{{\nu}_e}$=3.2 MeV (solid), electron anti-neutrinos T$_{\bar{\nu}_e}$=4.8 MeV  (dashed) and
for the other flavors 
T$_{\nu_x}$= 7.6 MeV (with $\nu_x=\nu_{\mu},\nu_{\tau},\bar{\nu}_{\mu},\bar{\nu}_{\tau}$) (dotted line).}
\label{fig:FD}
\end{figure}

We have calculated the neutrino evolution outside the supernova core using Eqs.(\ref{mixing}-\ref{wolfen2})
and determined the neutrino fluxes Eqs.(\ref{e:ratio}-\ref{e:lum}) and the electron
fraction (\ref{e:Yeeq0}-\ref{e:capture}). 
The numerical results we present are obtained
with a supernova density profile having a $1/r^{3}$ behavior  
(with the entropy per baryon, $S=70$ in units of
Boltzmann constant), 
that fits the numerical simulations shown in \cite{Balantekin:2004ug}.
The neutrino fluxes at the neutrinosphere are taken as Fermi-Dirac distributions with typical
temperatures of T$_{{\nu}_e}$=3.17 MeV,T$_{\bar{\nu}_e}$=4.75 MeV and
T$_{\nu_x}$= 7.56 MeV (with $\nu_x=\nu_{\mu},\nu_{\tau},\bar{\nu}_{\mu},\bar{\nu}_{\tau}$) (Figure \ref{fig:FD}) (the chemical potentials are assumed to be zero for simplicity).
The oscillation
parameters are fixed at the present best fit values \cite{Yao:2006px}, 
namely
$\Delta m^2_{12}= 8 \times 10^{-5}$eV$^2$, sin$^2 2\theta_{12}=0.83$ and
$\Delta m^2_{23}= 3 \times 10^{-3}$eV$^2$, sin$^2 2\theta_{23}=1$ for
the solar and atmospheric differences of the mass squares and
mixings, respectively. For the third still unknown neutrino mixing angle
$\theta_{13}$, we take either the present upper limit 
sin$^2 2\theta_{13}=0.19$ at 90 $\%$ C.L. (L) or  a very small value of 
sin$^2 2\theta_{13}=3 \times 10^{-4}$ (S)
 that might be attained at the future (third generation)  long-baseline
experiments \cite{Volpe:2006in}. Note that the value of $\theta_{13}$ determines the
adiabaticity of the first MSW resonance at high density
\cite{Dighe:1999bi,Engel:2002hg}, while $\theta_{12}$ governs the
second (adiabatic) one at low density. Since the sign of  the
atmospheric mixing is unknown, we consider both the normal (N) and
inverted (I) hierarchy. In the former (latter) case (anti)neutrinos
undergo the resonant conversion. We will denote results for the normal hierarchy and sin$^2 2\theta_{13}=0.19$
(N-H), inverted and sin$^2 2\theta_{13}=0.19$ (I-H), normal hierarchy and sin$^2 2\theta_{13}=$3. 10$^{-4}$
(N-S), inverted and sin$^2 2\theta_{13}=$3. 10$^{-4}$ (I-S). 

\begin{figure}[t]
\vspace{.6cm}
\includegraphics[scale=0.3,angle=0]{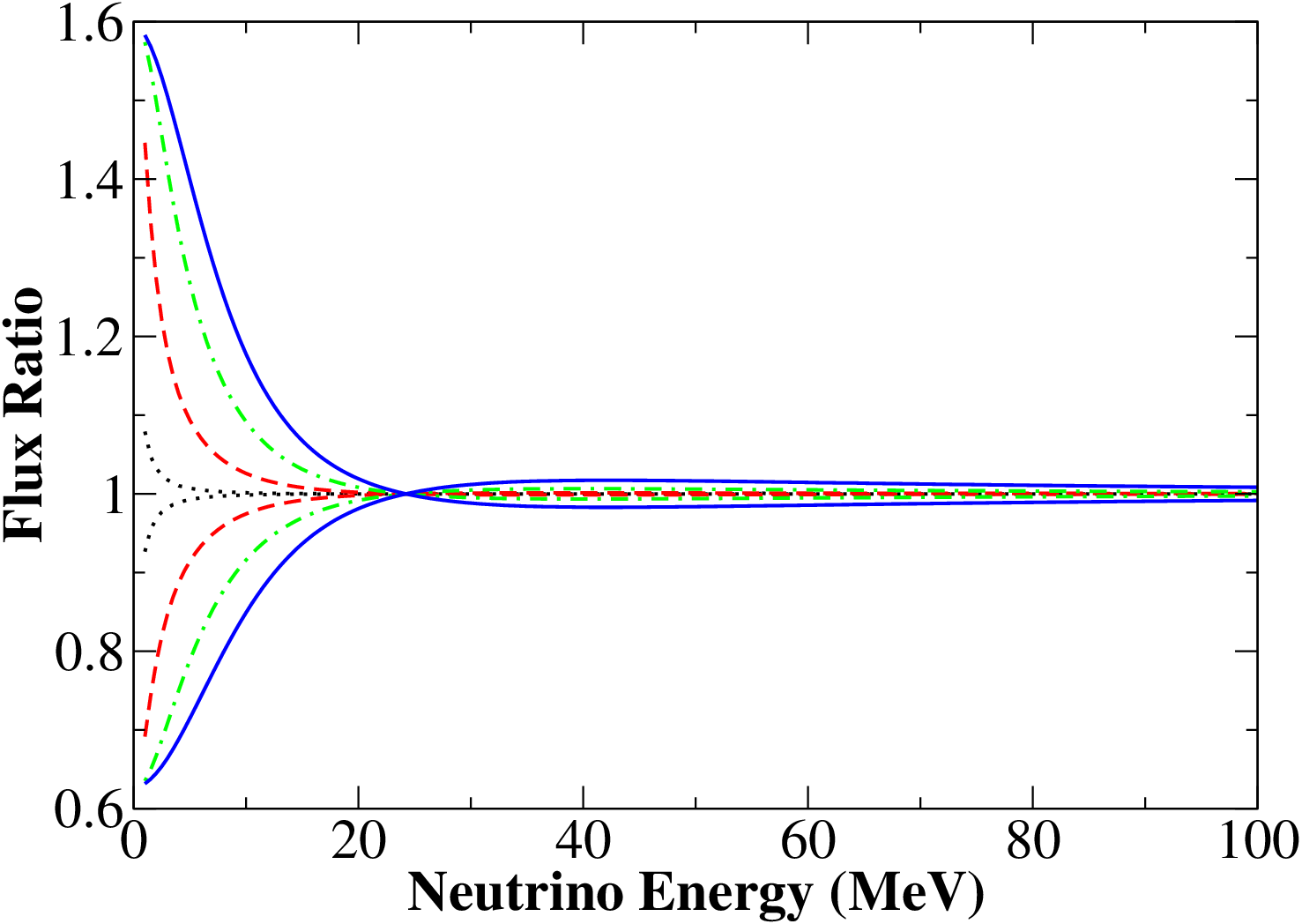}
\caption{$\bar{\nu}_{\mu}$ (lower curves) and $\bar{\nu}_{\tau}$ (upper curves) flux ratios for
a CP violating phase $\delta=180^{\circ}$ over $\delta= 0^{\circ}$ Eq.(\ref{e:ratio}), 
as a function of neutrino energy. Results at different distances from
the neutron-star surface are shown, namely 250 km (dotted), 500 km (dashed),
750 km (dot-dashed) and 1000 km (solid line). The curves correspond to the normal
hierarchy and sin$^2 2\theta_{13}=0.19$.  \label{fig:numuCP}}
\end{figure}

\begin{figure}[t]
\vspace{.6cm}
\centerline{\includegraphics[scale=0.3,angle=0]{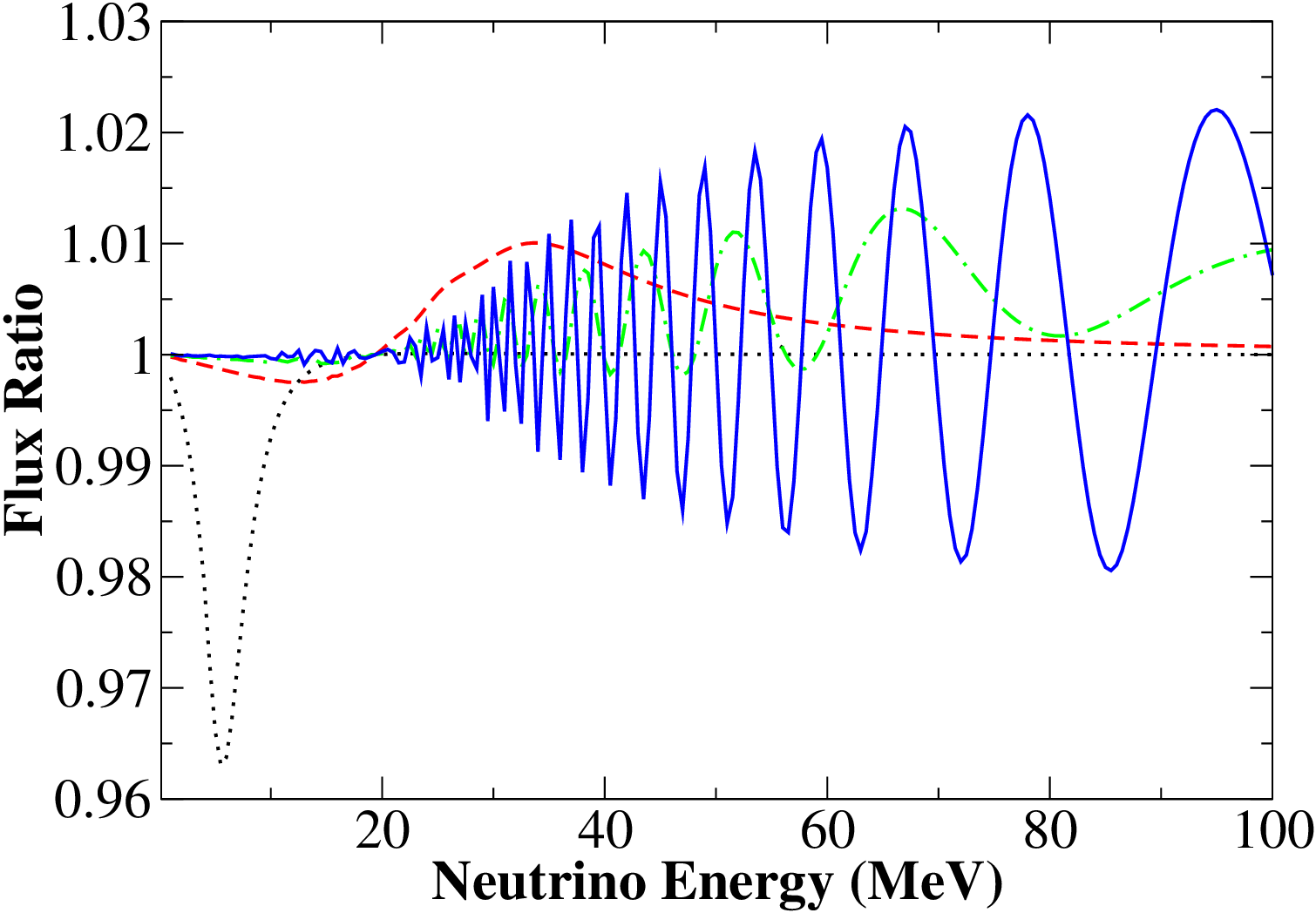}\hspace{.2cm}
\includegraphics[scale=0.3,angle=0]{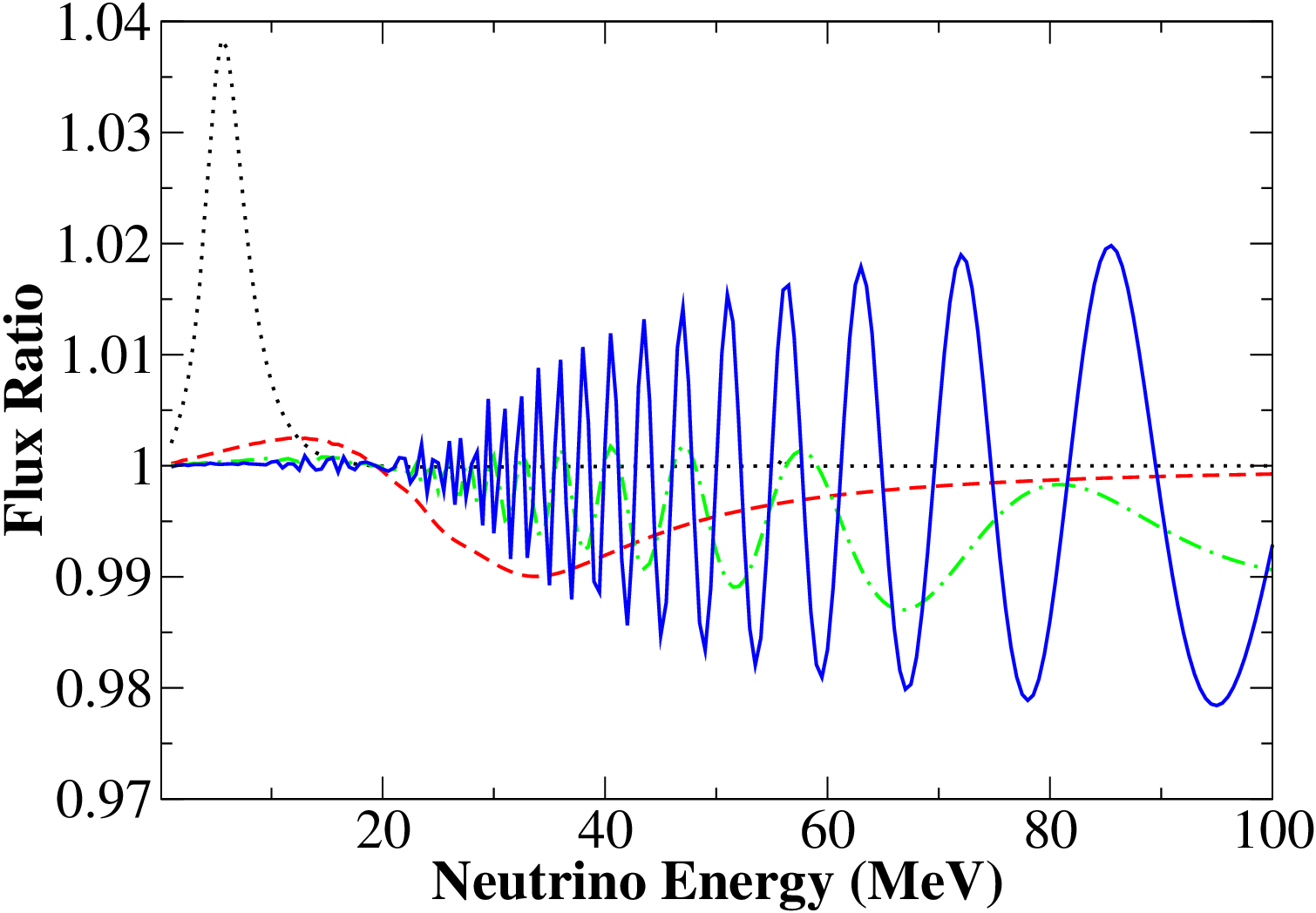}}
\caption{Same as Fig.\ref{fig:numuCP} for $\nu_{\mu}$ (left) and
$\nu_{\tau}$ (right) fluxes.\label{fig:nutauCP}}
\end{figure}

\begin{figure}[t]
\vspace{.6cm}
\centerline{\includegraphics[scale=0.3,angle=0]{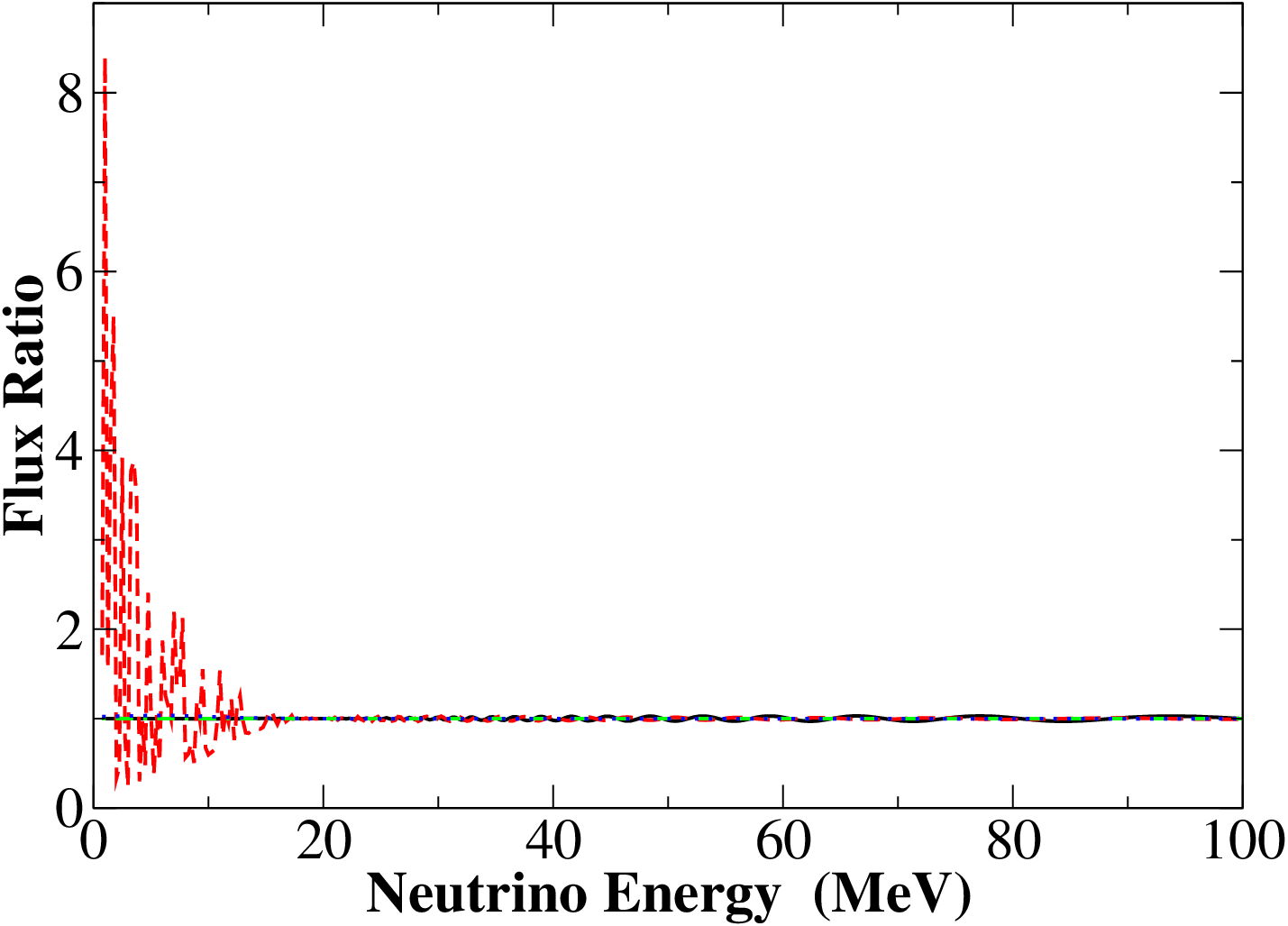}\hspace{.2cm}
\includegraphics[scale=0.3,angle=0]{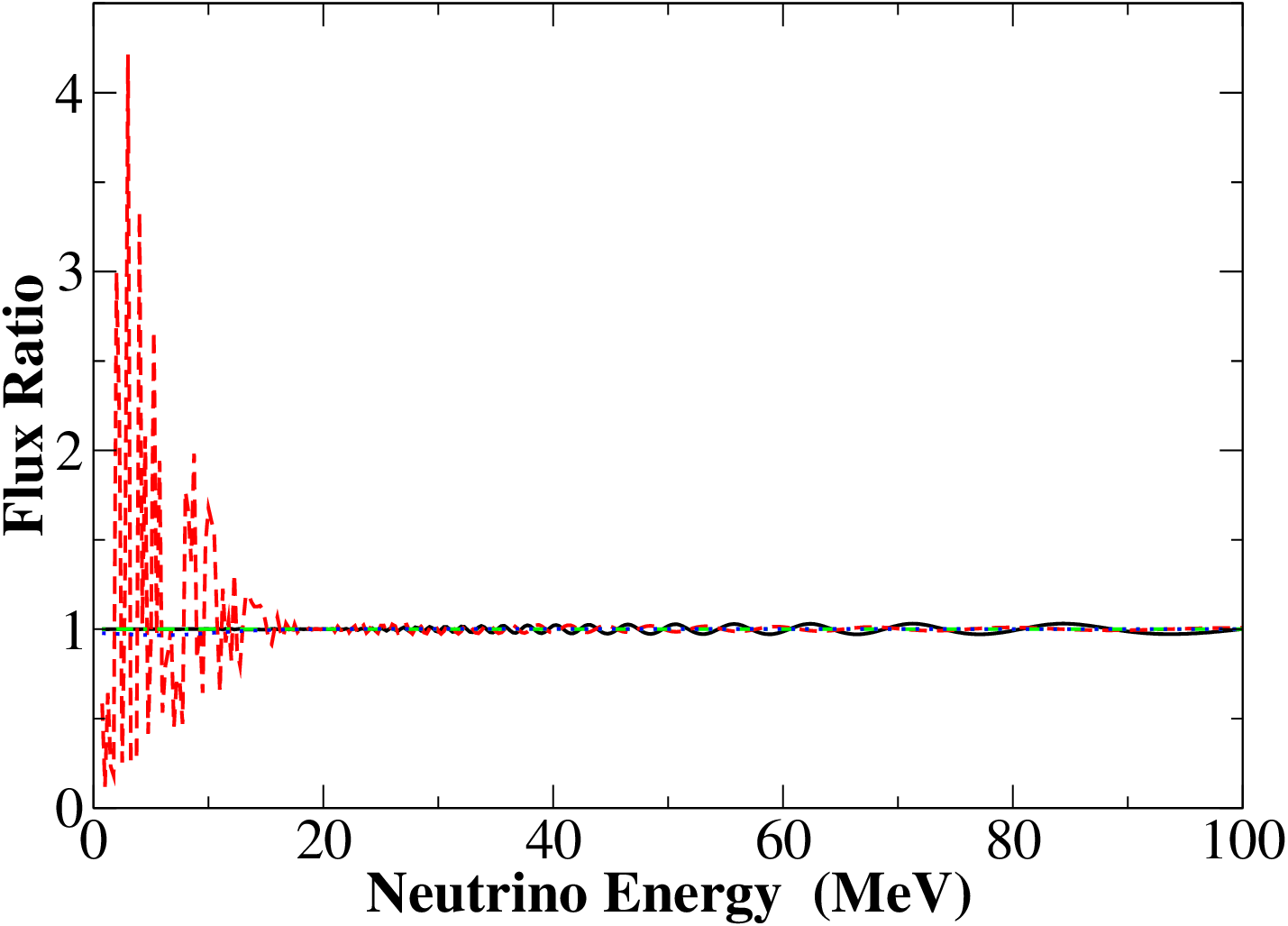}}
\caption{Ratio of the $\nu_{\mu}$ (left) and $\nu_{\tau}$ (right) fluxes for
$\delta=180^{\circ}$ over $\delta=0^{\circ}$ at a distance of 1000 km
from the neutron-star surface. The curves correspond to N-L (solid), N-S (dashed), I-L (dot-dashed), I-S (dotted). \label{fig:nmuratio}}
\end{figure}

Figures \ref{fig:numuCP} and  \ref{fig:nutauCP} show the $\bar{\nu}_{\mu},\bar{\nu}_{\tau}$ and
$\nu_{\mu},\nu_{\tau}$ flux ratios
Eq.(\ref{e:ratio})  for $\delta=180^{\circ}$ over for
$\delta=0^{\circ}$.  
One can see that large effects, up to 60 $\%$ are present for
low neutrino energies in the anti-neutrino case; while smaller effects,
of the order of a few percent, appear in the neutrino case. 
The effect of a non-zero delta over the $\nu_{\mu},\nu_{\tau}$ fluxes as a function of
neutrino energy is shown in Figure \ref{fig:nmuratio} at a distance of 1000 km. We see that an increase as large as
a factor of 8 (4) can be seen at low energies in the $\nu_{\mu}$ ($\nu_{\tau}$) spectra.
A similar behavior is found in the anti-neutrino case.

In most of supernova
simulations, the $\nu_{\mu}$ and  $\nu_{\tau}$ luminosities
are approximately equal,  because these particles interact via neutral current
only, at the low energies possible at supernova\footnote{Note however that, even at the level
of the Standard Model, some differences can arise for example from loop corrections
\cite{Botella:1986wy}.}.  
Since the $\nu_e,\bar{\nu}_e$ appearance probabilities are
independent of $\delta$ and as long as the $\nu_{\mu}$ and  $\nu_{\tau}$
luminosities are taken to be equal,  using Eqs.(\ref{sq2}) and (\ref{e:flux}) one can show that
 the $\nu_e$ and $\bar{\nu}_e$ fluxes are independent of the CP violating phase. 
Practically all the literature concerning the neutrino evolution in core-collapse supernovae 
ignore the Dirac phase, for simplicity. 
Our results justify this assumption if
such calculations make the hypothesis that the 
$\nu_{\mu}$ ($\bar{\nu}_{\mu}$) and $ \nu_{\tau}$ ($\bar{\nu}_{\tau}$) 
luminosities are equal and neglect the $V_{\mu,\tau}$. 

\begin{figure}[h]
\vspace{.6cm}
\centerline{\includegraphics[scale=0.29,angle=0]{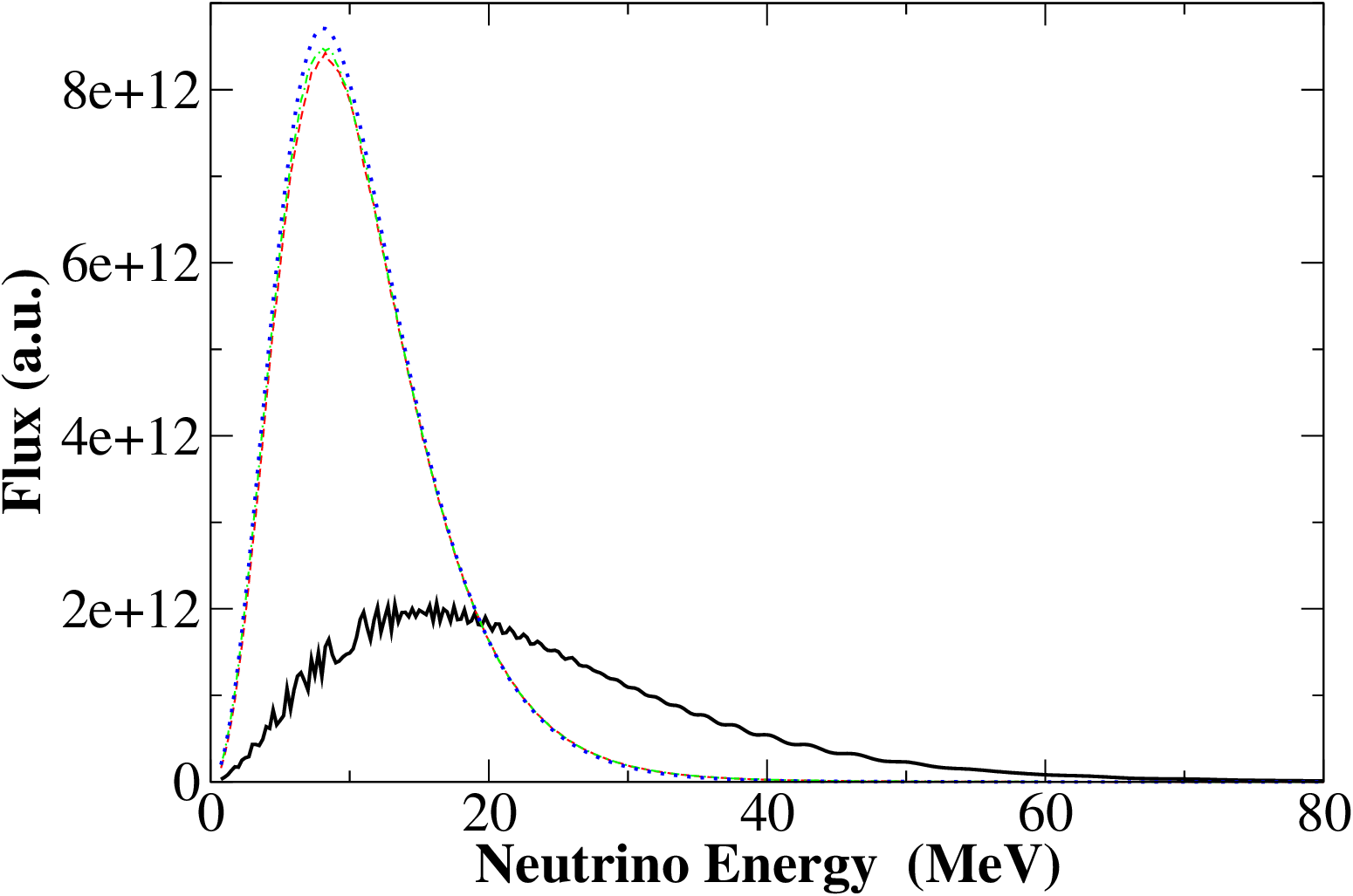}\hspace{.2cm}
\includegraphics[scale=0.29,angle=0]{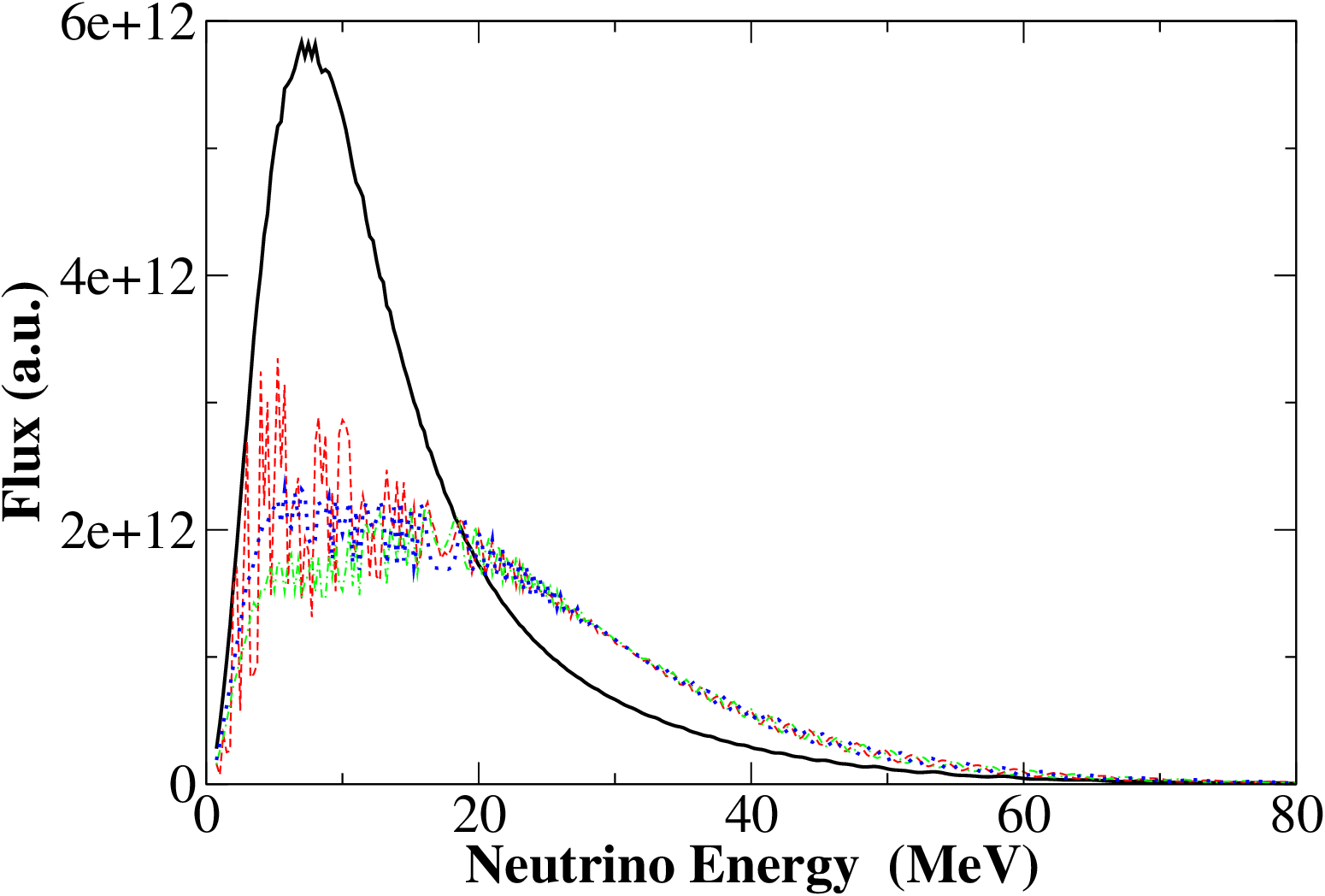}}
\caption{Electron (left) and muon (right) neutrino fluxes Eq.(\ref{e:flux}) at 1000
km from the neutron star surface, N-L (solid), N-S (dashed), I-L (dot-dashed), I-S (dotted).
In the N-L case, the first resonance is adiabatic and
the Fermi-Dirac $\nu_e$ distributions at the
neutrinospere (Fig.\ref{fig:FD}) are completely swapped with $\nu_x$.
The situation is reversed for $\nu_{\mu}$.
These results are obtained by fixing  T$_{\nu_{\tau}}$ larger
than T$_{\nu_{\mu}}$ by 1 MeV, as an example of the difference that could
be induced by the presence of flavor-changing interactions in the
neutrinosphere (see text).\label{fig:evolvedflux}}
\end{figure}

\begin{figure}[h]
\vspace{.6cm}
\centerline{\includegraphics[scale=0.3,angle=0]{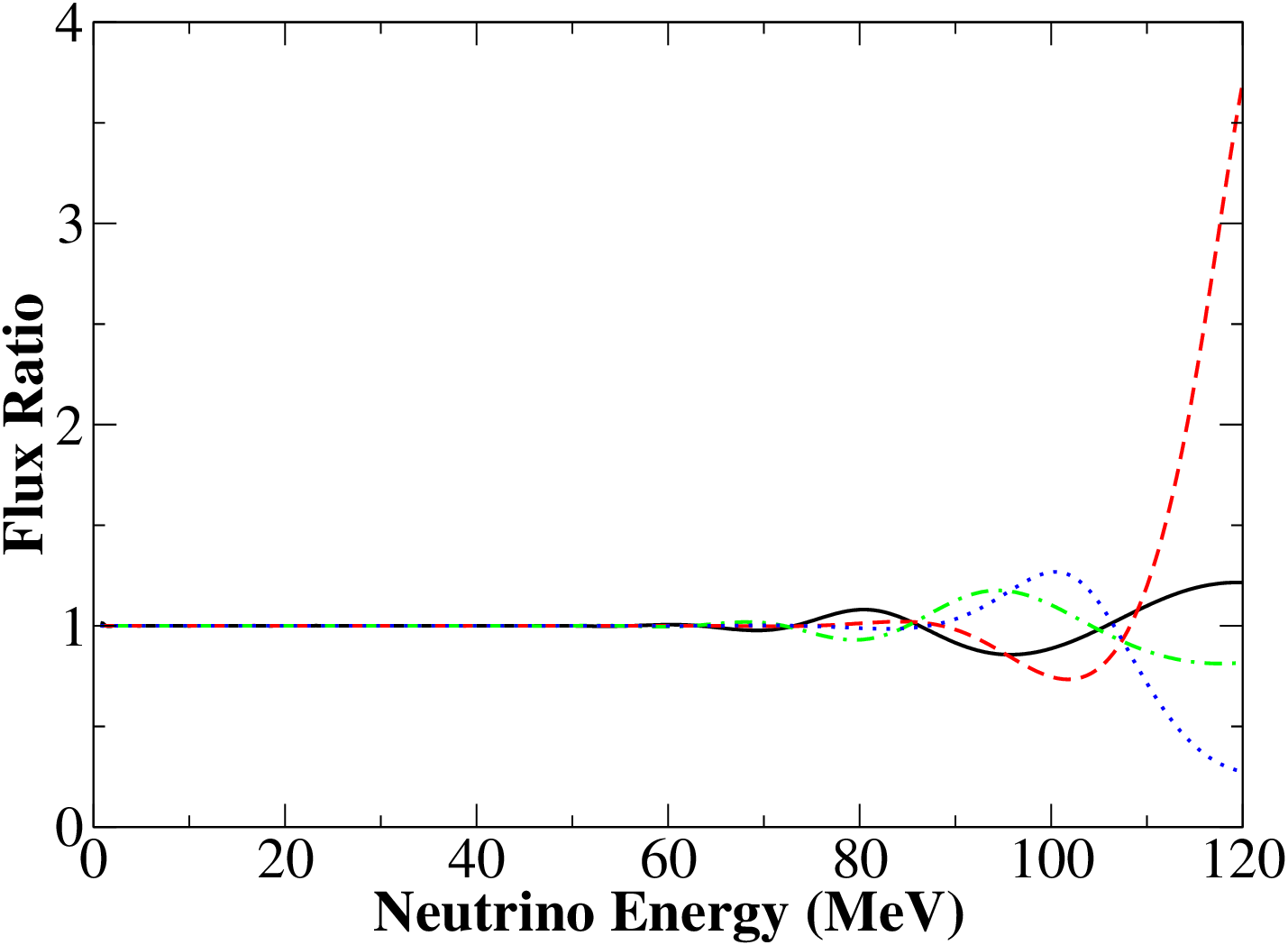}\hspace{.2cm}
\includegraphics[scale=0.3,angle=0]{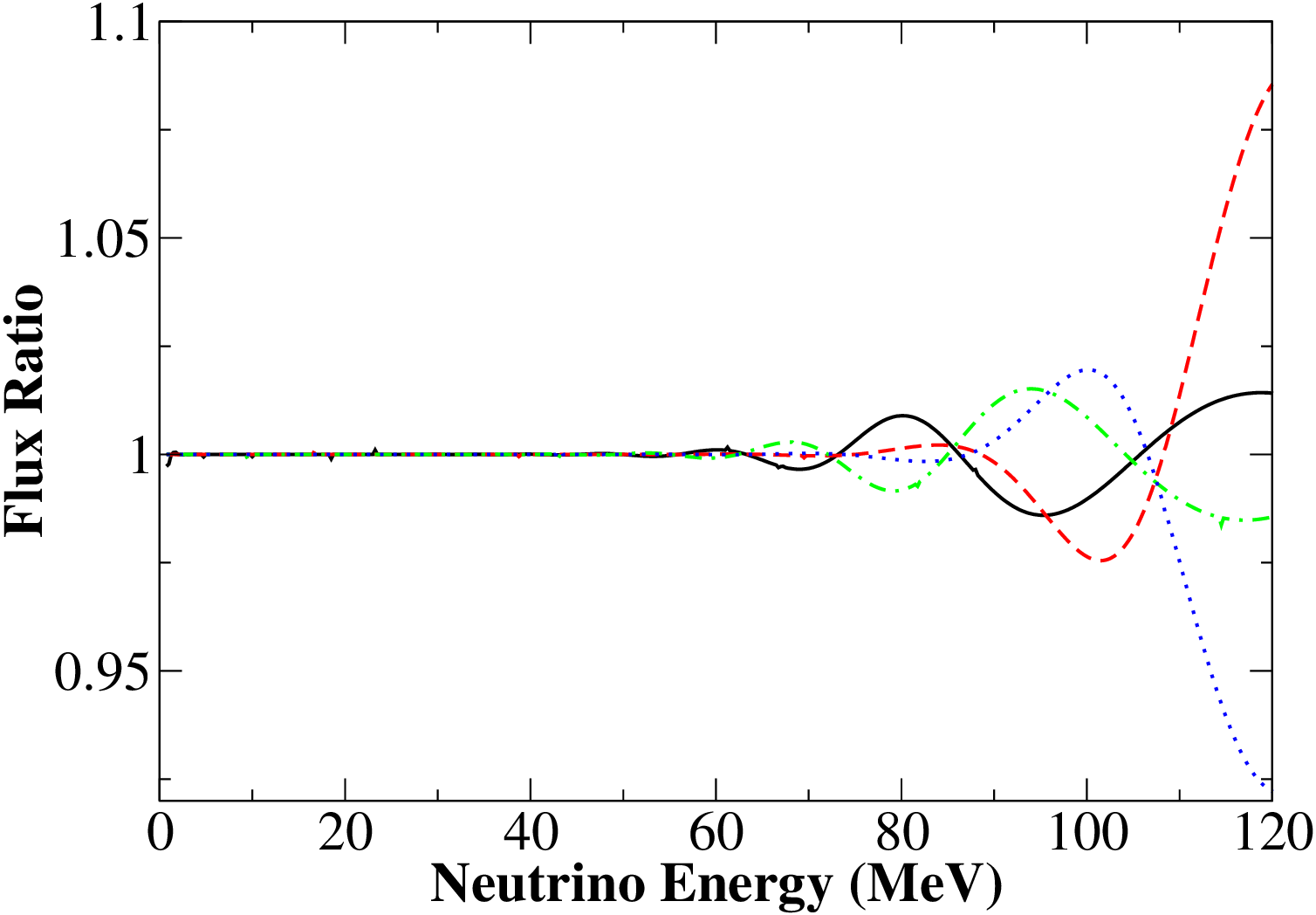}}
\caption{Ratios of the $\nu_e$ flux $\delta=180^{\circ}$ over for
$\delta=0^{\circ}$ at 200 km from the neutron star surface, obtained by taking 
$L^0_{\nu_{\tau}}=1.1~L^0_{\nu_{\mu}}$ (right) or
T$_{\nu_{\tau}}$= 8.06 MeV and T$_{\nu_{\mu}}$= 7.06 MeV (left)
(see text).
The curves correspond to N-L (solid), N-S (dashed), I-L (dot-dashed), I-S (dotted).
\label{fig:nueratioTLdif200}}
\end{figure}

\begin{figure}[h]
\vspace{.6cm}
\centerline{\includegraphics[scale=0.3,angle=0]{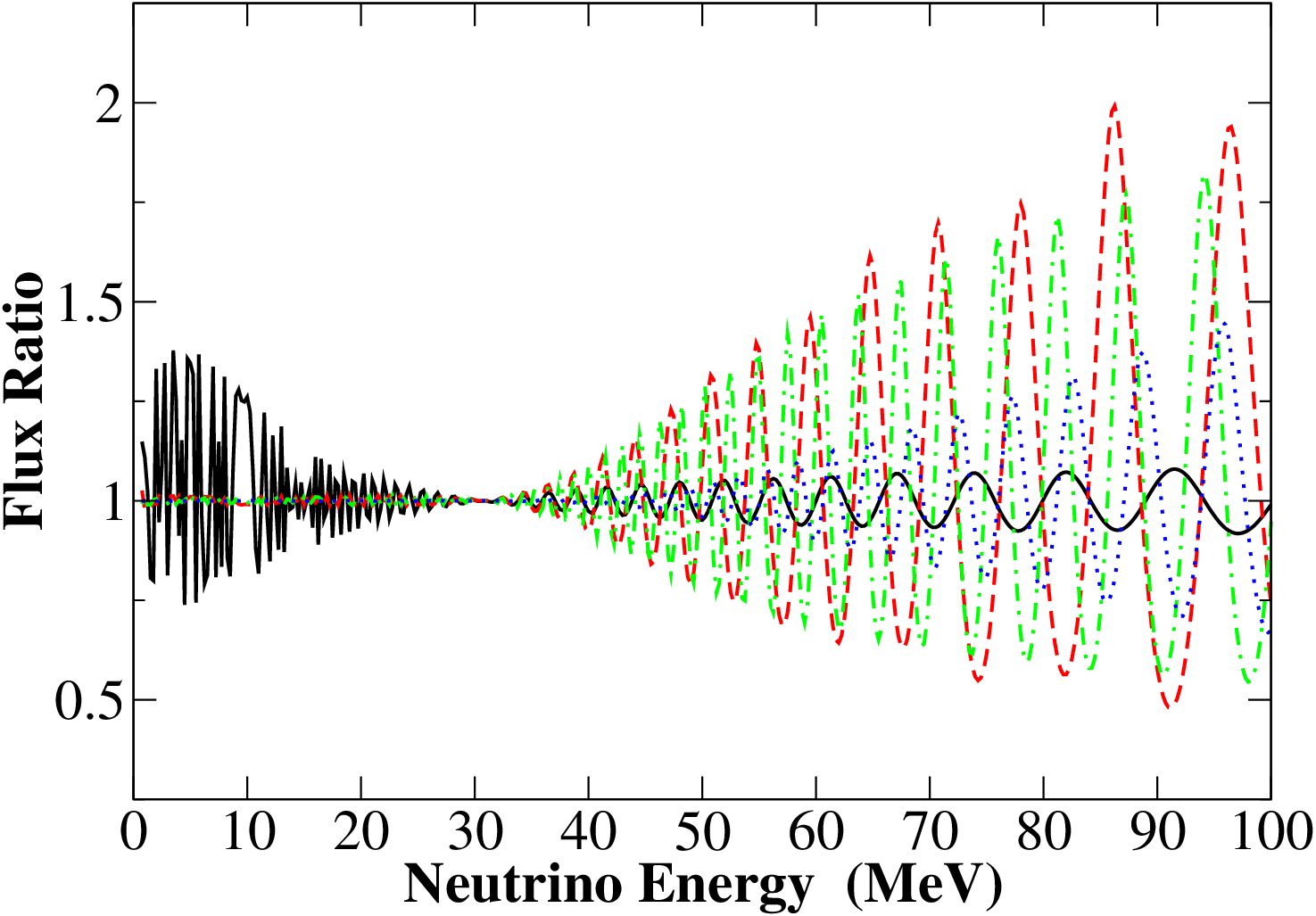}\hspace{.2cm}
\includegraphics[scale=0.3,angle=0]{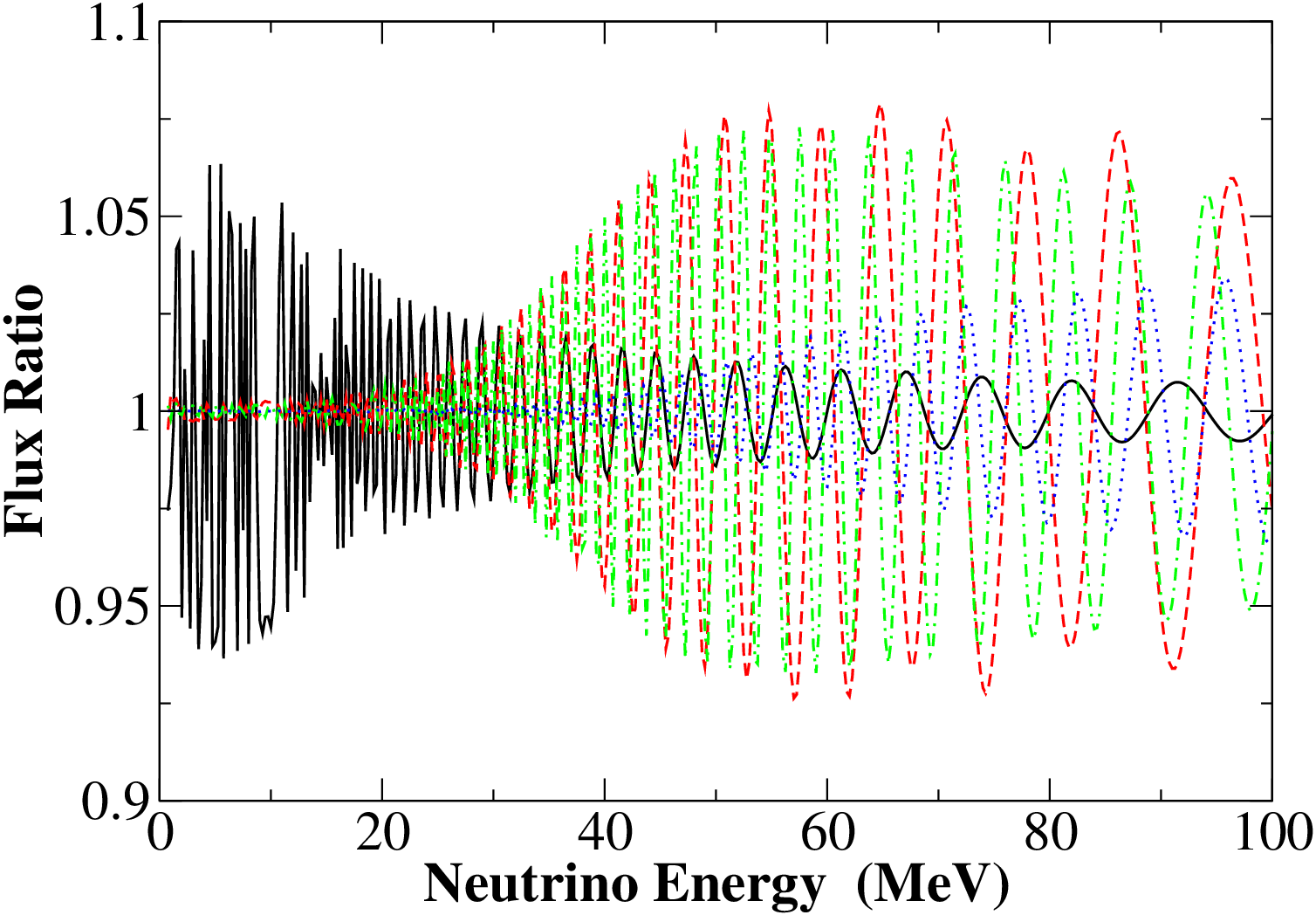}}
\caption{Same as Fig.\ref{fig:nueratioTLdif200} but at 1000 km. 
\label{fig:nueratioTLdif1000}}
\end{figure}

\begin{figure}[t]
\vspace{.6cm}
\centerline{\includegraphics[scale=0.3,angle=0]{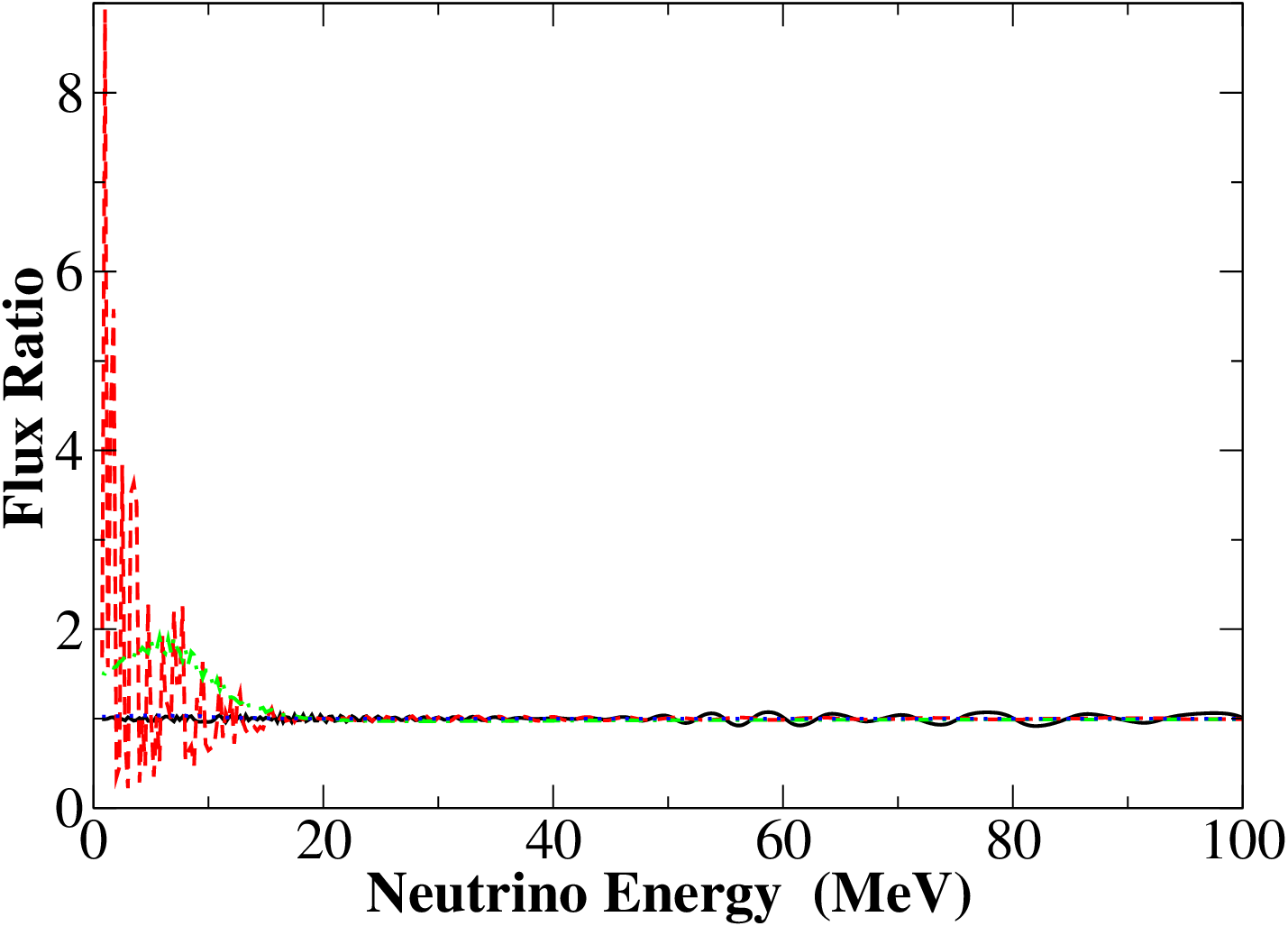}\hspace{.2cm}
\includegraphics[scale=0.3,angle=0]{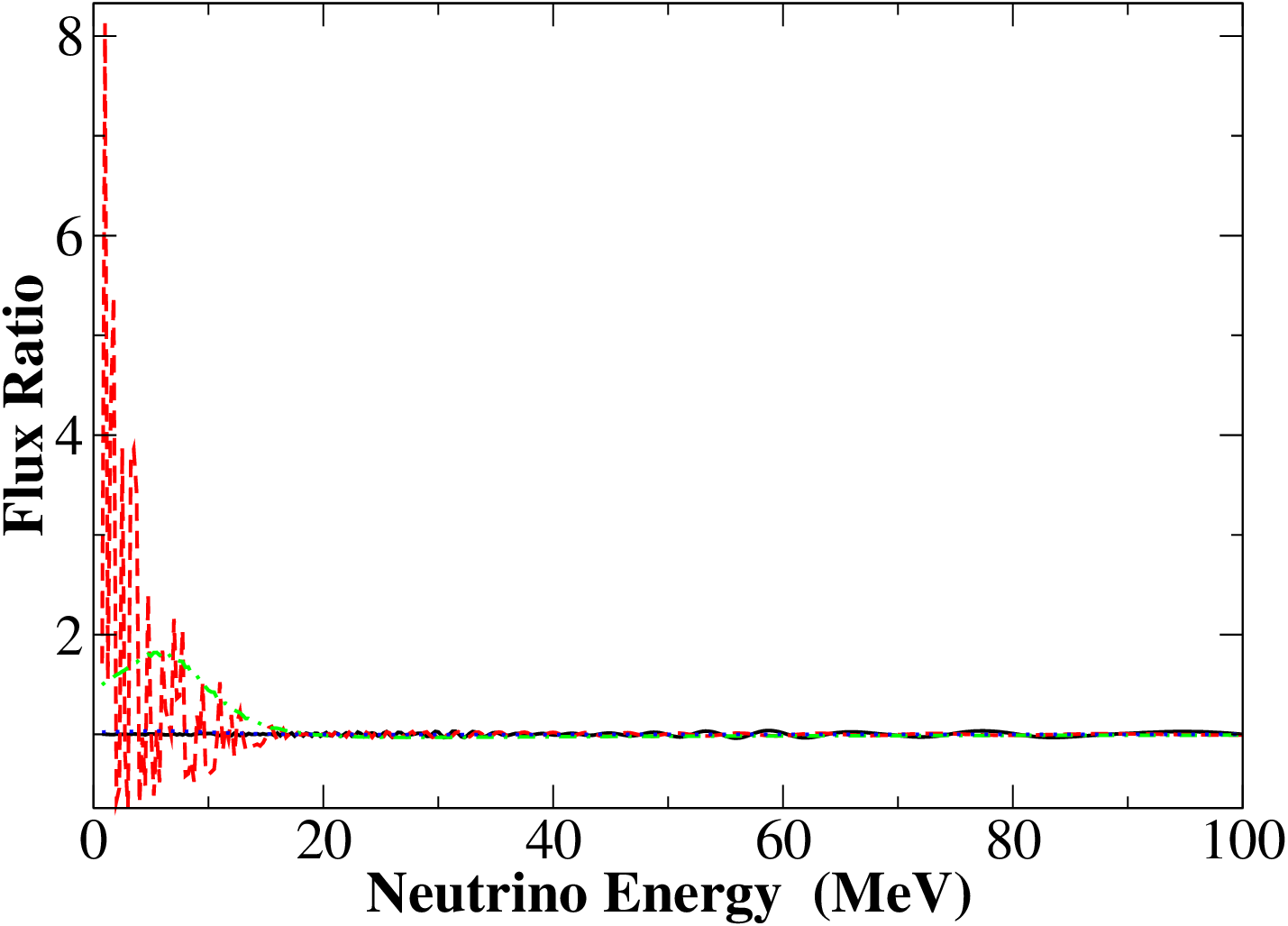}}
\caption{Same as Fig.\ref{fig:nueratioTLdif1000} but for the $\nu_{\mu}$ flux ratios.\label{fig:numuratioTLdif}}
\end{figure}

On the other hand, the situation is different 
if the muon and tau neutrino fluxes are different at the neutrinosphere 
either because of the corrections
within the Standard Model and/or because of
physics beyond the Standard Model, such as flavor changing
interactions \cite{Amanik:2004vm} which might populate the
$\nu_{\mu},\nu_{\tau}$ fluxes differently and differentiate their
temperatures at decoupling. 
Our aim is to show the CP violating effects in this case. 
We have explored various differences between the $\nu_{\mu},\nu_{\tau}$ luminosities. We present here for example results corresponding to
the $\nu_{\mu},\nu_{\tau}$ total luminosities Eq.(\ref{e:flux}) 
different by 10 $\%$, e.g. $L^0_{\nu_{\tau}}=1.1~L^0_{\nu_{\mu}}$ or 
T$_{\nu_{\tau}}$= 8.06 MeV while T$_{\nu_{\mu}}$= 7.06 MeV.  
Figure \ref{fig:evolvedflux} presents as an example 
the evolved $\nu_e,\nu_{\mu}$ neutrino fluxes, at 1000 km from the neutron star surface, 
when T$_{\nu_{\mu}} \ne T_{\nu_{\tau}}$.
The different curves show results for the two hierarchies 
and the two values of $\theta_{13}$. Similarly to the case where T$_{\nu_{\mu}}=T_{\nu_{\tau}}$,
while for the N-L case the first resonance is adiabatic and the electron neutrinos get a hotter spectrum,
for all other cases the spectra keep very close to the Fermi-Dirac distributions (Figure \ref{fig:FD}).
The situation is obviously reversed for the muon neutrino flux.
Figures \ref{fig:nueratioTLdif200},
\ref{fig:nueratioTLdif1000} and 
\ref{fig:numuratioTLdif} show the ratios of the $\nu_e$ and $\nu_{\mu}$ fluxes
for a non-zero over a zero delta phase, as a function of neutrino
energy, at different distances from the neutron star surface. 
One can see that effects up to a factor of 2-4 
on the $\nu_e$ and up to 10 $\%$ on $\nu_{\mu}$ are present.
A similar behavior is found for the $\bar{\nu}_{e}$ and $\nu_{\tau}$ fluxes.

The behavior of the flux ratios shown in Figure \ref{fig:nueratioTLdif1000} is easy to 
understand. From Eqs. (\ref{mue}) and (\ref{taue}) one can write
\begin{eqnarray}
\label{oscillation}
\phi_{\nu_e} (\delta) &=& \phi_{\nu_e} (\delta =0) + \sin 2\theta_{23} 
\sin \frac{\delta}{2} 
(L_{\nu_\tau} - L_{\nu_\mu}) \\ &\times& 
\left[ \sin 2\theta_{23} \sin \frac{\delta}{2} 
(|B_{\mu e}|^2 - |B_{\tau e}|^2) + \left[ \left(
\cos 2\theta_{23} \sin \frac{\delta}{2} 
- i \cos \frac{\delta}{2} \right)
(B_{\mu e} B^*_{\tau e}) + {\rm h.c.} \right] \right] \nonumber 
\end{eqnarray}
Clearly the ratios calculated in these figures would be identity 
at the value of the energy where $\nu_{\mu}$ and $\nu_{\tau}$ spectra 
would cross (i.e., $L_{\nu_\tau} = L_{\nu_\mu}$). Away from this energy 
one expects an oscillatory behavior due to the additional terms in Eq. 
(\ref{oscillation}) as the figure indicates. 
Note that even for $\delta=0$ the neutrino fluxes could also exhibit an oscillatory behavior. 
Concerning Fig. \ref{fig:numuratioTLdif}, one can see that
the effects due to $\delta \ne 0$ and 
induced by taking different temperatures or luminosities are small, compared
to the case with $\delta \ne 0$ only (Figure \ref{fig:nmuratio}). 

\begin{figure}[h]
\vspace{.6cm}
\centerline{\includegraphics[scale=0.3,angle=0]{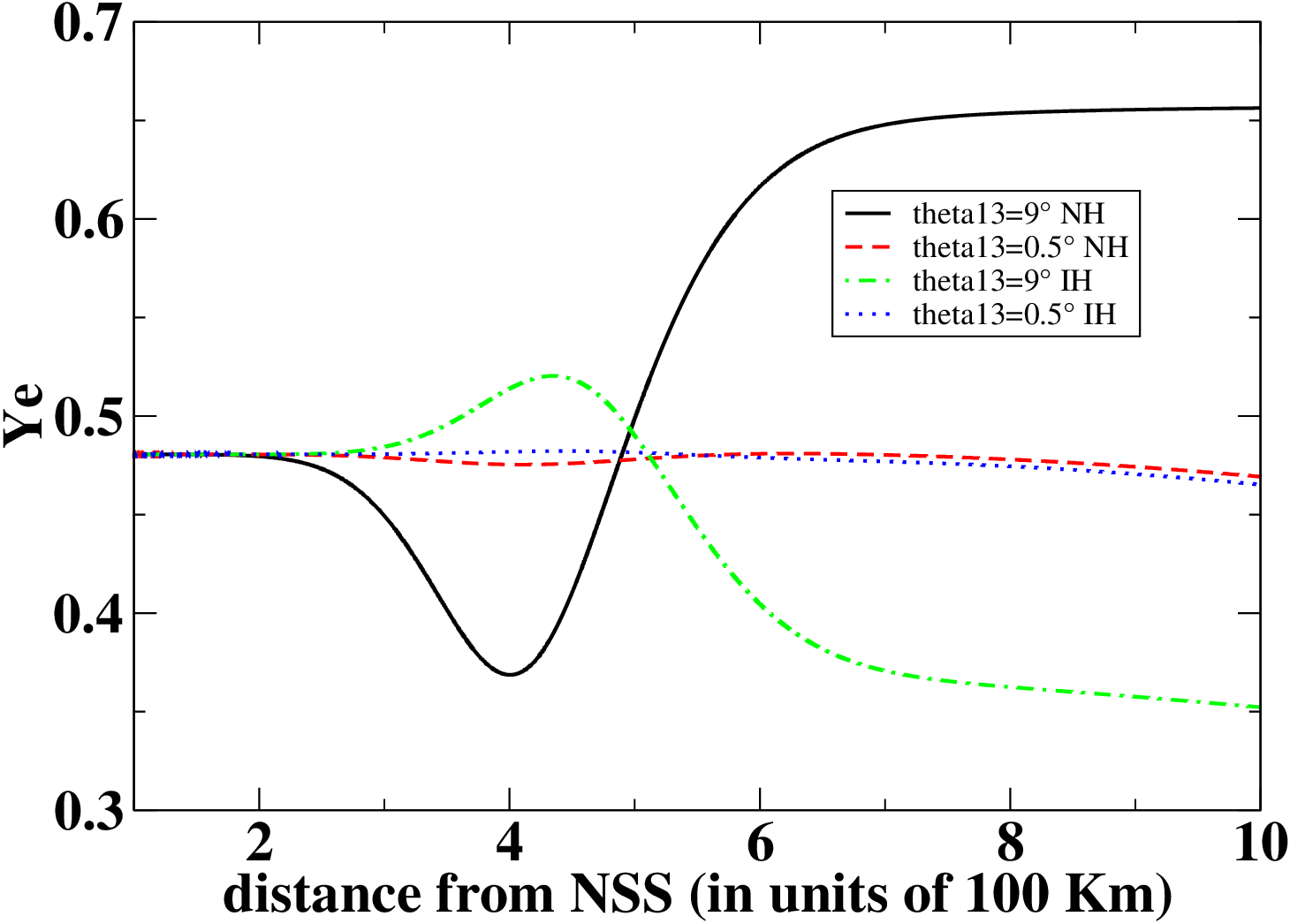}\hspace{.2cm}
\includegraphics[scale=0.3,angle=0]{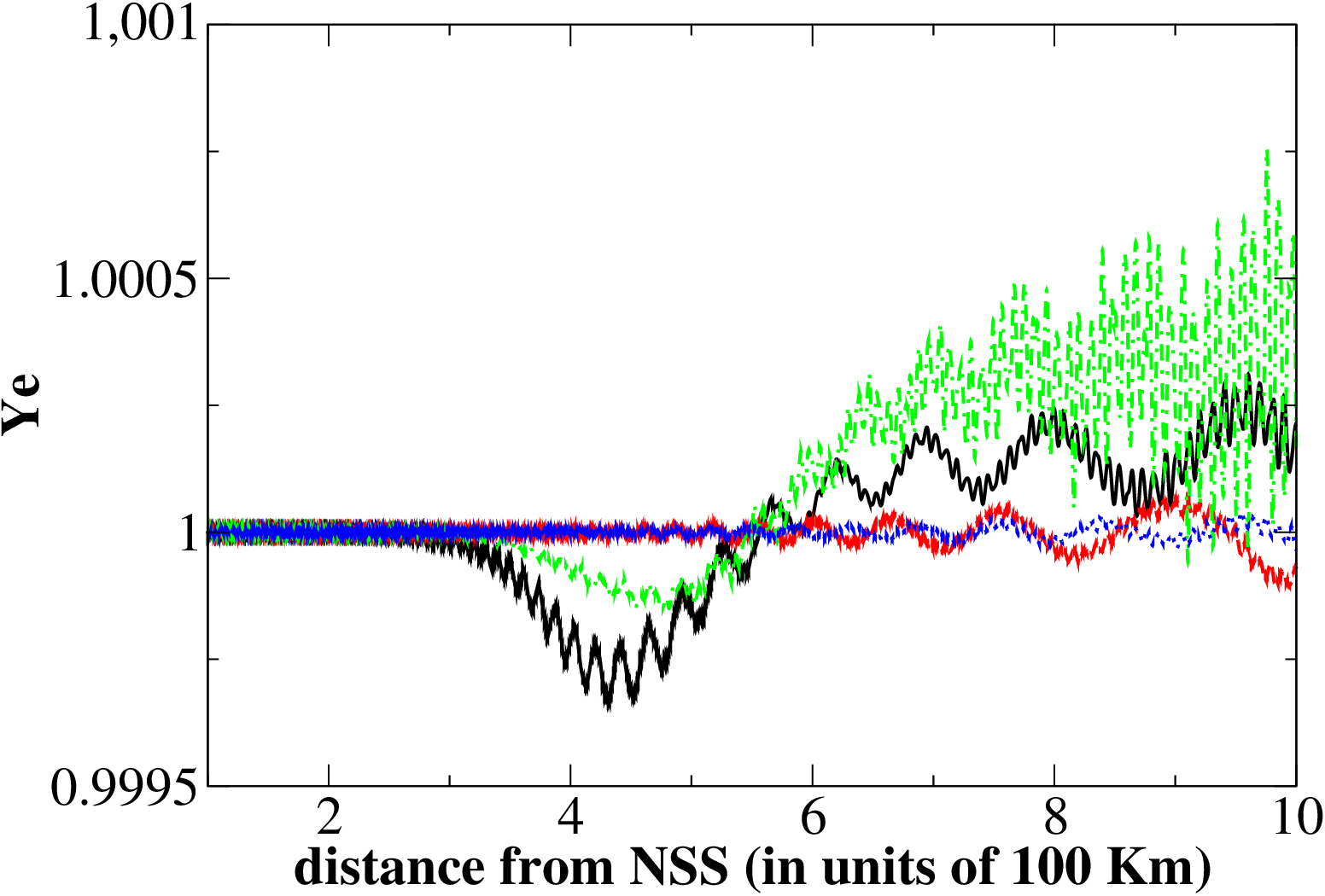}}
\caption{Electron fraction for $\delta=0$ (left) and
ratios of the electron fraction (right) for $\delta=180^{\circ}$
compared to $\delta=0^{\circ}$,
as a function of the distance from the neutron-star surface. The
initial $\nu_{\mu},\nu_{\tau}$ fluxes have temperatures which differs
by 1 MeV (see text).  The results correspond to the normal hierarchy
and sin$^2 2\theta_{13}=0.19$. \label{fig:YediffT}}
\end{figure}

Figure \ref{fig:YediffT} shows results on the electron fraction $Y_e$.
Note that if $\delta \ne 0$ there are no CP violation effects on $Y_e$ since this quantity
depends on the electron neutrino and anti-neutrino fluxes only Eqs.(\ref{e:Yeeq0},\ref{e:capture}).
Our results show that 
the effects due to $\delta \ne 0$ 
are small (of the order of $0.1 \%$) in all the studied cases with different muon and tau total luminosities and/or temperatures.

\begin{figure}[t]
\vspace{.6cm}
\centerline{\includegraphics[scale=0.3,angle=0]{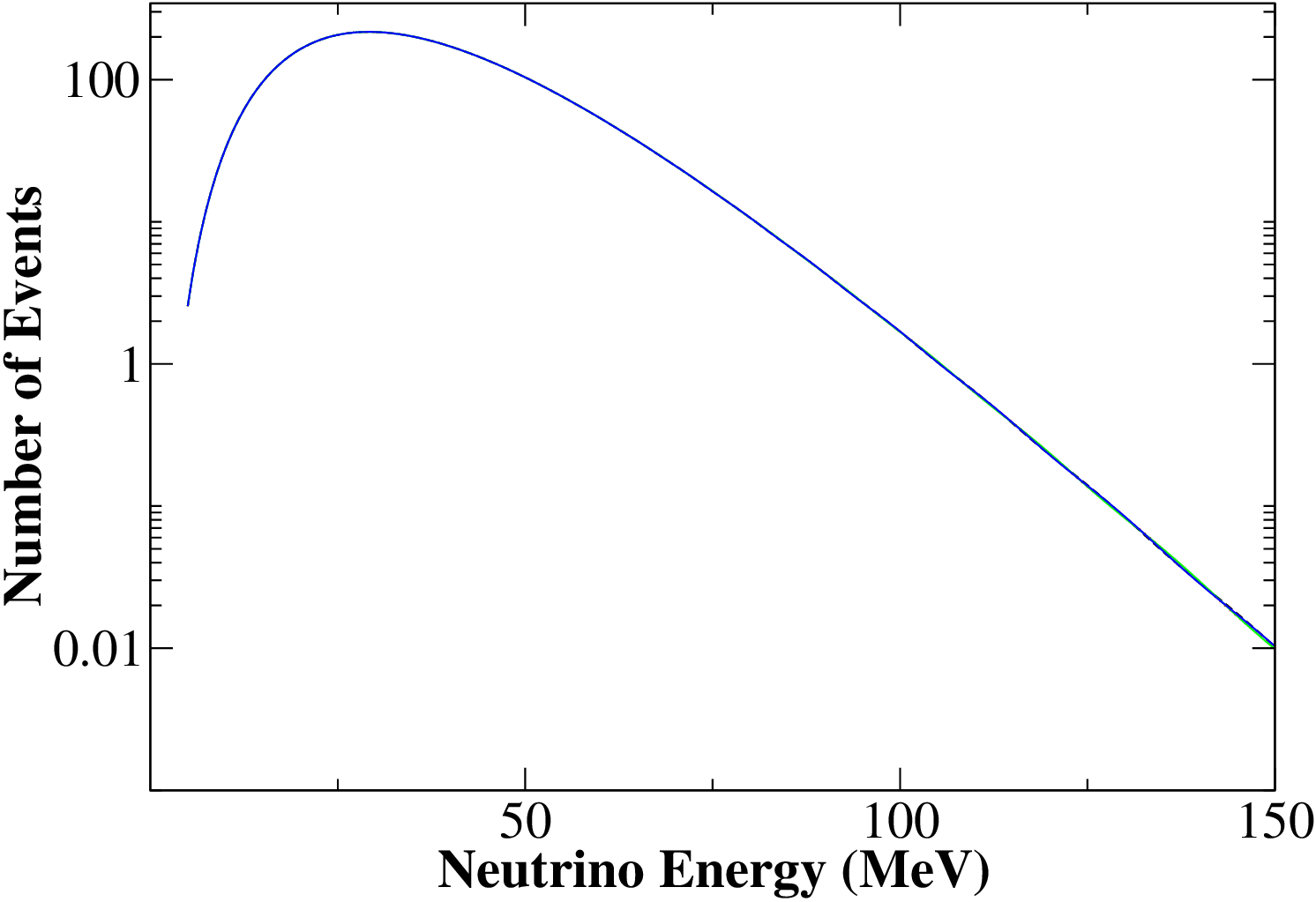}\hspace{.2cm}
\includegraphics[scale=0.3,angle=0]{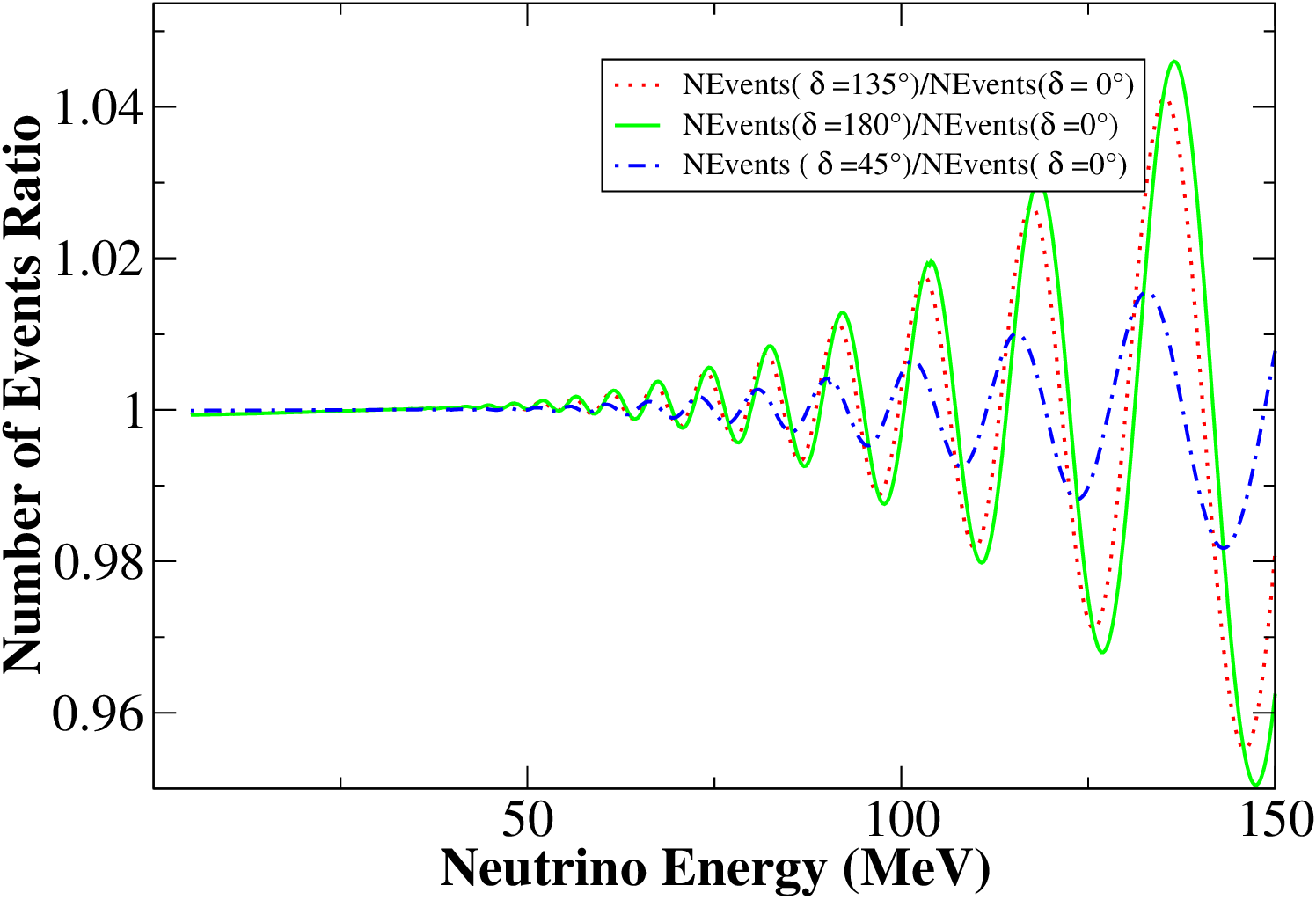}}
\caption{Number of events associated to $\bar{\nu}_e + p \rightarrow n + e^+$ 
from a possible future supernova explosion at 10 kpc in a detector like 
Super-Kamiokande (22.5 ktons). These results are obtained for 
inverted hierarchy and large third mixing angle.}\label{fig:events}
\end{figure}

Finally, we discuss the effects induced by the CP violating phase
$\delta$ on the supernova neutrino signal in a terrestrial
observatory. 
Figure \ref{fig:events}
presents the expected number of events associated to electron anti-neutrino scattering on protons
 for different $\delta$ values. 
This is calculated by convoluting the fluxes from Eq.(\ref{e:flux}-\ref{e:lum}) by the relevant
anti-neutrino proton cross section \cite{Balantekin:2006ga}.
A water \v{C}erenkov detector such as Super-Kamiokande (22.5 Ktons)
is considered as an example. We assume 100 \% efficiency. Note that the neutral current signal which is sensitive to all fluxes turns out to be $\delta$ independent as well, as can be shown by adding the three fluxes 
Eq.(\ref{e:flux}).
One cas see that $\delta$
phase induces small modifications up to 5 $\%$  in the number of events, as a function of neutrino
energy, and of the order of $2. 10^{-4}$ on the total number of events. In fact, for a supernova 
at 10 kpc, we get for inverted hierarchy and large third neutrino mixing angle
7836.1 for $\delta=45^{\circ}$,
7837.0  for $\delta=135^{\circ}$,
7837.2 for $\delta=180^{\circ}$;
while it is 7835.9 for $\delta=0^{\circ}$. 
These results are obtained with muon and tau neutrino
fluxes having difference temperatures. 
Similar conclusion are drawn if we take different luminosities.
For normal hierarchy and large $\theta_{13}$,
effects of the same order are found while for small  $\theta_{13}$ and inverted/normal
hierarchy the effecs become as small as $10^{-5}$ .

\section{Conclusions}
\noindent
In this work we have analyzed possible effects induced by the CP violating Dirac phase in a dense environment such as the core-collapse supernovae. 
Our major result are that in matter:
i) significant effects are found on the muon and tau neutrino fluxes for a non-zero
CP violating phase;
ii) important effects are also found on the electron (anti)neutrino fluxes if
the $\nu_{\mu}$ and $\nu_{\tau}$ neutrino fluxes differ at the neutrinosphere.
On the other hand 
the usual assumption of ignoring the CP violating phase 
made in the literature is justified if contributions from physics beyond the Standard Model
is small and the $\nu_{\mu}$ and $\nu_{\tau}$ fluxes are equal at emission.
We have calculated the events in an observatory on earth and shown that 
effects at the level of $5 \%$ are present on the number of events as a function
of neutrino energy.

Recent calculations
have shown that the inclusion of neutrino-neutrino interaction introduces 
new features in the neutrino propagation in supernovae. A detailed study of the neutrino evolution with the 
CP violating phase, the neutrino-neutrino interaction as well
as loop induced neutrino refractive indices will be the object of further work.

%%%%%-------------------------------------------------------------%
\section*{ACKNOWLEDGMENTS}
%-----------------------------------------------------------------%
J.Gava and C.Volpe thank Takashi Yoshida for useful discussions.
The authors acknowledge the CNRS-Etats Unis 2006 and 2007 grants which
have been used during the completion of this work.  J.G. and
C.V. acknowledge the support from the ANR-05-JCJC-0023 "Non standard
neutrino  properties and their impact in astrophysics and cosmology".
This work was also supported in part by the U.S. National Science
Foundation Grant No.\ PHY-0555231 and in part by the University of
Wisconsin Research Committee with funds granted by the Wisconsin
Alumni Research Foundation.

\end{document}